\documentclass[letterpaper, 10 pt, conference]{ieeeconf}  

\usepackage{graphicx}
\usepackage{subfigure}
\usepackage{amssymb}
\usepackage{amsmath}
\newtheorem{remark}{Remark}
\newtheorem{problem}{Problem}

\newtheorem{proposition}{Proposition}

\graphicspath{{figures/}}

\IEEEoverridecommandlockouts                              
\title{\LARGE \bf
A Formal Verification Approach to the Design of \\ Synthetic Gene Networks
}

\author{Boyan Yordanov and Calin Belta
\thanks{This work was supported by NSF grant CNS-0834260. B. Yordanov ({\tt\small yordanov@microsoft.com}) was with the Department of Mechanical Engineering, Boston University, Boston MA and is now with Microsoft Research, Cambridge UK. C. Belta ({\tt\small cbelta@bu.edu}) is with the Department of Mechanical Engineering, Boston University, Boston%
}
}

\begin{document}

\maketitle
\thispagestyle{empty}
\pagestyle{empty}


\begin{abstract}
The design of genetic networks with specific functions is one of the major goals of synthetic biology. However, constructing biological devices that work ``as required" remains challenging, while the cost of uncovering flawed designs experimentally is large. To address this issue, we propose a fully automated framework that allows the correctness of synthetic gene networks to be formally verified {\em in silico} from rich, high level functional specifications.

Given a device, we automatically construct a mathematical model from experimental data characterizing the parts it is composed of. The specific model structure guarantees that all experimental observations are captured and allows us to construct finite abstractions through polyhedral operations. The correctness of the model with respect to temporal logic specifications can then be verified automatically using methods inspired by model checking.

Overall, our procedure is conservative but it can filter through a large number of potential device designs and select few that satisfy the specification to be implemented and tested further experimentally. Illustrative examples of the application of our methods to the design of simple synthetic gene networks are included.
\end{abstract}


\section{INTRODUCTION}

Synthetic biology is an emerging field that focuses on the rational design of biological systems. A number of {\em biological devices} - engineered gene networks that function as switches, oscillators, counters, logic gates, etc. have been designed, implemented, and validated experimentally, demonstrating the feasibility of the approach (see \cite{Purnick2009} for a review). However, success stories have largely been the result of extensive manual effort, application of various modeling and analysis techniques, and trial and error experimentation. As the field matures, real world applications are sought requiring a systematic approach that enables the implementation of complicated designs into functionally correct devices with less experimental work.

One such approach has been enabled by biological standards~\cite{Knight2003} and (online) part libraries~\cite{MITparts} and involves the modular design and construction of devices from {\em biological parts} - genetic sequences known to function as promoters, ribosome binding sites, coding sequences, etc. This, in turn, has allowed the development of Bio-Design Automation (BDA) platforms, such as \texttt{Clotho}~\cite{Densmore2009}, which provide an environment where online libraries can be accessed, devices can be designed using a graphical interface and checked against rules of correct assembly, while the construction process is completed automatically by liquid handling robots. Even so, current BDA platforms asses potential devices only in terms of their assembly feasibility and not based on their correctness with respect to specifications of required function.

Designing biological devices that work ``as required" remains challenging and, to minimize costly experimentation, it is usually approached through modeling (see \cite{deJong02} for a review of modeling formalisms). A realistic model is needed to guide design efforts but such models are often hard to analyze. In addition, estimating model parameters may require extensive experimental data which is rarely available, although characterizations resulting in biological part data sheets~\cite{Canton2008} are currently ongoing. Besides selecting a realistic yet analytically tractable model, specifying the required device behavior in a formalism that is both general and allows for automatic analysis procedures is a separate challenge. In this paper, we propose a fully automated framework for {\em in silico} verification of synthetic gene networks from rich, high level specifications expressed as temporal logic formulas.

Temporal logics~\cite{Clarke99} are customarily used for specifying the correctness of digital circuits and computer programs. Due to their expressivity and resemblance to natural language they have gained popularity in other areas including the specification and analysis of qualitative behavior of genetic networks~\cite{Antoniotti03,Batt2005,Batt08}. There also exist off-the-shelf algorithms for verifying the correctness of a finite state system for a temporal logic specification (model-checking). However, such finite models are usually too simple to capture the dynamics of genetic networks with the detail necessary for design applications.

In our previous work~\cite{Yordanov2010} we used piecewise affine (PWA) systems as models of gene networks~\cite{Sontag81}. Such systems are globally complex and can approximate nonlinear dynamics with arbitrary accuracy~\cite{Lin1992}, which makes them realistic models. They are also locally simple, which allowed us to analyze them formally from temporal logic specifications through a procedure based on the construction and refinement of finite abstractions through polyhedral operations~\cite{Yordanov2010} and  model-checking~\cite{Clarke99}. In this paper, we use a class of models that is inspired by PWA systems but is more general. To account for the variability due to experimental conditions and the uncertainty inherent in biological systems we allow model parameters to vary in some ranges. We develop a procedure for the automatic construction of such models from part characterization data with the guarantee that all experimental observations can be reproduced by the identified model. We also extend our methods from \cite{Yordanov2010} and integrate them with our model identification procedure, which leads to a fully automatic framework for specifying and verifying the correctness of genetic networks constructed from parts. Our approach can be used both to verify individual device designs or to automatically explore the space of potential device designs that can be constructed from characterized parts, available from libraries.

In terms of analysis, our method is related to tools such as the \texttt{Genetic Network Analyzer (GNA)}~\cite{deJong03} and \texttt{RoVerGeNe}~\cite{Batt07}, which study biological systems using temporal logic specifications but usually focus on the analysis of separate devices for which a model is available. Instead, our procedure can explore different devices constructed from a set of parts, while models are derived automatically from part characterization data. In that respect, our approach resembles methods such as \cite{Little2010} that focus on the generation and analysis of analog and mixed-signal circuit models from simulation traces. Unlike other gene network modeling approaches, we do not enforce sigmoidal (Hill) regulation functions but construct models which capture all experimental observations and resemble uncertain parameter PWA systems. Our procedure is therefore also related to methods for the identification of PWA systems (see~\cite{Juloski05} for a review). Such tools address the threshold reconstruction problem more rigorously but only identify fixed parameter models and require device experimental data which is not usually available during design. This motivates the development of our procedure for constructing device models from part characterization data.

Automatic construction of device models from kinetic parameters of parts has been implemented in \texttt{Asmparts}~\cite{Rodrigo2007}, \texttt{SynBioSS}~\cite{Hill2008}, \texttt{GEC}~\cite{Pedersen09} and \texttt{GenoCAD}~\cite{Cai2009}. These tools can be used to search for devices with specific functions by numerically simulating potential device models and assessing the ``goodness" of the resulting trajectories but face three main challenges. First, it is assumed that parameters of individual parts are known which is not always the case. In this paper, we only assume that protein degradation rates are known, motivated by the fact that such information is often available from literature or can be predicted computationally~\cite{Gasteiger2005}. However, we compute expression rates from experimental data, which makes our approach easier to apply in practice. Second, the models resulting from other procedures are hard to analyze and, therefore, numerical simulation is used to asses device behavior. This can be unfeasible for large state spaces and, in general, does not lead to any formal guarantees. We construct models which are rich enough to capture all experimental observations but can also be analyzed formally using model-checking based methods, which we have developed previously~\cite{Yordanov2010} but extend in this paper. Finally, no general formalism for specifying required device behavior is provided in the outlined procedures and only solutions for specific cases are considered. We formalize high level specifications in linear temporal logic which is both rich ({\em i.e.} it captures many properties of interest) and ``user-friendly" ({\em i.e.} it resembles natural language).

The remainder of this paper is organized as follows. In Sec. \ref{sec:problem} we formulate the main problem and present an overview of our approach. In Sec. \ref{sec:identification} we discuss the automatic construction of models from part characterization data. We focus on the formal analysis of those models through the construction of finite abstractions in Sec. \ref{sec:analysis}. In Sec. \ref{sec:case_study} we present illustrative examples of the application of our framework to the design of a bistable synthetic gene network. We conclude with final remarks and directions of future work in Sec. \ref{sec:conclusion}.

Throughout the rest of the paper we use the following notation. Given a set $S$ we use $|S|$ and $2^S$ to denote the cardinality and the powerset (the set of subsets) of $S$, respectively. For a set $S \subset \mathbb{R}^N$ and a scalar $\lambda \in \mathbb{R}$, we use $\lambda S$ to denote the set of elements from $S$ multiplied by $\lambda$. Given sets $S$ and $S'$ we denote their Minkowski (set) sum by $S+S'$. Given polytope $X$, we denote the set of vertices of $X$ by $\mathcal{V}(X)$ and their convex hull as $X = hull(\{v \in \mathcal{V}(X)\})$.

\section{PROBLEM FORMULATION} \label{sec:problem}
In this section we formulate the problem of verifying the correctness of a gene network (biological device) from high level specifications. We start by discussing our simplified view of the biochemistry involved in gene expression, the parts we consider as basic building blocks of all devices and the experimental data that we assume is available.

\begin{figure}[t]
\centering
\includegraphics[scale=0.63]{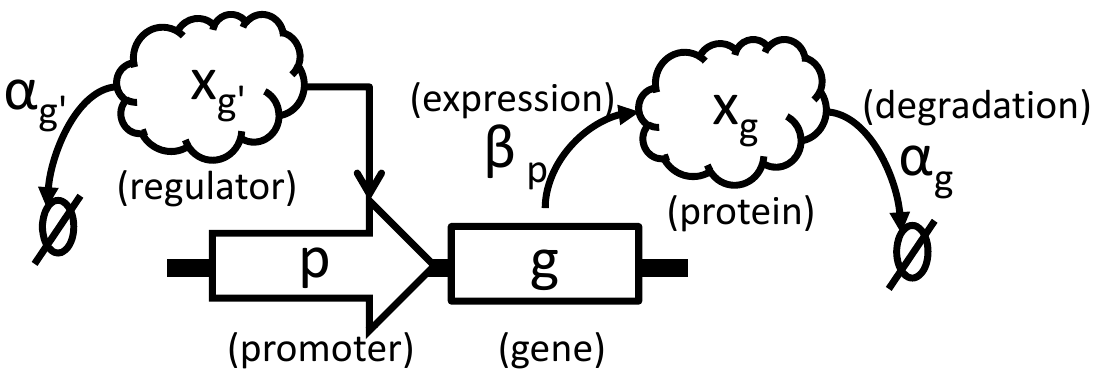}
\caption{In the simplified expression mechanism we consider, a gene $g$ is expressed from promoter $p$ at rate $\beta_p$ to produce protein, whose concentration is denoted as $x_g$. The protein degrades at rate $\alpha_g$. The promoter might be regulated by protein $x_{g'}$.}\label{fig:pic1}
\vspace{-0.25in}
\end{figure}

We consider only two basic types of biological parts - sequences of DNA that either function as {\em promoters} or code for proteins (we refer to such sequences as {\em genes}). This is the minimal set of parts required to define the interactions in gene networks but the methods we subsequently develop can be extended easily for more detailed formulations. We assume that each gene codes for a single protein which degrades at a certain rate and whose concentrations can be measured directly in experiments. We treat protein production as a single step process, which is sufficient to capture transcriptional regulation (see Fig. \ref{fig:pic1}).

For a protein to be produced, its corresponding gene must be expressed, which requires placing it after a promoter (we assume that other sequences required for correct expression such as a ribosome binding site are already contained within a gene). The simplest device we consider contains a single promoter and expresses a single gene to produce a single protein (Fig. \ref{fig:pic1}). By placing multiple genes on the same promoter and including additional promoters, more complicated devices can be assembled. We assume that, in a device, a gene is expressed from a single promoter - such assembly constraints are handled by platforms such as \texttt{Clotho}~\cite{Densmore2009}.

We differentiate between {\em constitutive} and {\em regulated} promoters. A protein is always produced if its gene is expressed from a constitutive promoter ({\em i.e.} the promoter is always ``on"), while expression from a regulated promoter varies, depending on the concentrations of proteins or chemicals (inducers), called {\em regulators}. In general, a promoter can be regulated by several regulators but, for simplicity of presentation in this paper we consider only the case of a single regulator per promoter, although our methods can also be extended for the more general case.

\begin{figure}[t]
\subfigure[Constitutive promoter]{\includegraphics[scale=0.45]{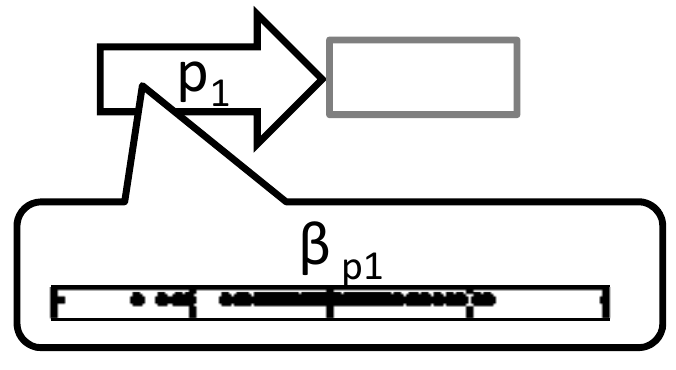}\label{fig:pic2}}
\subfigure[Regulated promoter]{\includegraphics[scale=0.50]{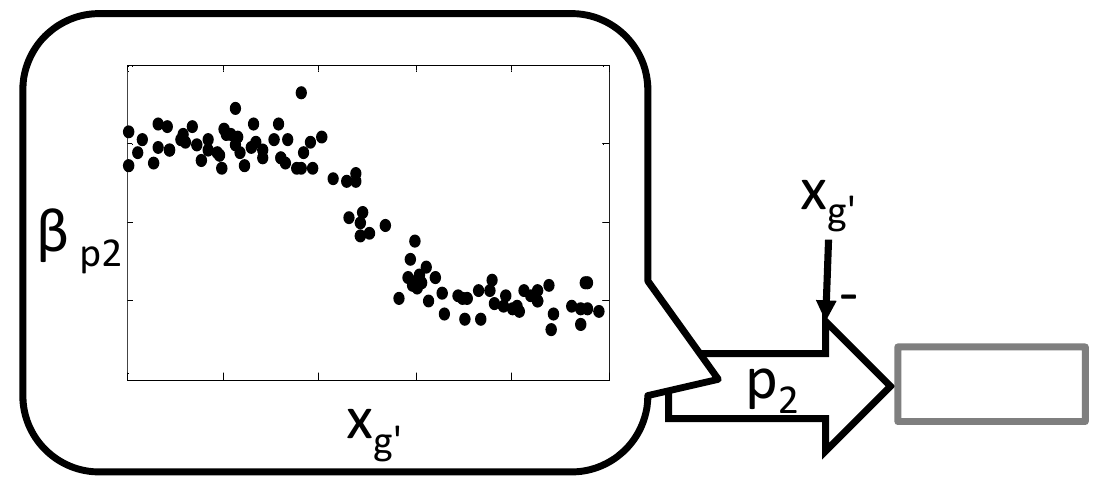}\label{fig:pic3}}
\caption{\subref{fig:pic2} Experimental observations in the form of a range of values of the rate of expression $\beta_{p1}$ characterize the constitutive promoter $p_1$. \subref{fig:pic3} The regulated promoter $p_2$ is characterized by experimental observations of the rate of expression $\beta_{p2}$ at different concentrations of the regulator (in this case, repressor) $x_{g'}$.}\label{fig:formulation_pics}
\vspace{-0.25in}
\end{figure}

We consider only devices built from {\em characterized parts} - genes and promoters for which experimental data indicative of their performance is available. A gene is characterized by the degradation rate (or equivalently the half-life) of the protein it codes for, which we assume is a fixed and known value. Protein degradation rates are often available from literature or can be predicted computationally~\cite{Gasteiger2005}. A promoter is characterized by a rate of expression, which we assume is the same for all genes expressed from it. However, because of variability in experimental conditions and the inherent uncertainty of biological systems, we assume that the rate of expression from a promoter varies in a certain range. For a constitutive promoter, the characterization data is simply a set of experimentally measured expression rates (Fig. \ref{fig:pic2}), while for a regulated promoter, we assume that experimental measurements of the expression rate at different concentrations of the regulator are available (Fig. \ref{fig:pic3}). Measuring expression rates directly can be challenging and such data is usually obtained by simultaneously measuring the concentration of a regulator and a gene expressed from the regulated promoter~\cite{Rosenfeld2005}. In Sec. \ref{sec:identification} we provide a procedure for converting such measurements to the expression rates data shown in Fig. \ref{fig:formulation_pics}, which we assume is available for all characterized promoters.

Given a device, we are interested in studying the dynamics of the concentrations of proteins expressed from its genes. Let $G$ denote the set of genes where $N = |G|$ is the device size. We use $x_g$ to denote the concentration of the protein expressed from gene $g \in G$, which is bounded in a physiologically relevant range $x_g^{min} \leq x_g \leq x_g^{max}$. The hyper-rectangle
\begin{equation}\label{eqn:state_space}
\mathcal{X} = [x_{g_1}^{min},x_{g_1}^{max}] \times \ldots \times [x_{g_N}^{min},x_{g_N}^{max}]
\end{equation}
is the feasible state space of the device, where each $x \in \mathcal{X}$ is a vector of the concentrations of all proteins $x_g, g \in G$. Given an initial state $x(0) \in \mathcal{X}$ the concentrations of species from $G$ evolve over time and produce an infinite sequence $x(0),x(1),\ldots$ called a {\em trajectory}, where $x(k) \in \mathcal{X}$ is the state at step $k=1,\ldots$.

We define a set of atomic propositions $\Pi$ as a set of linear inequalities
\begin{equation}\label{eqn:ineq}
\Pi = \{\pi_i,i=1,\ldots,K\}, \pi_i = \{x \in \mathcal{X}\;|\; c_i^Tx+d_i\leq 0\}.
\end{equation}
In other words, each atomic proposition $\pi_i$ partitions the feasible space $\mathcal{X}$ into a satisfying and violating subset for $\pi_i$. Given a state $x \in \mathcal{X}$ we write $x \vDash \pi_i$ if and only if $c_i^Tx+d_i\leq 0$ ({\em i.e} $x$ satisfies $\pi_i$). A trajectory $x(0),x(1),\ldots$ produces an infinite word $w(0),w(1),\ldots$ where $w(k) = \{\pi \in \Pi\;|\; x(k) \vDash \pi\}$ is the set of propositions satisfied at step $k$.

To specify temporal logic properties of trajectories of the system we use Linear Temporal Logic~\cite{Clarke99}. Informally, LTL formulas over $\Pi$ are inductively defined by using the standard Boolean operators ({\em e.g.,} $\neg$ (negation), $\vee$ (disjunction), $\wedge$ (conjunction)) and temporal operators, which include $\bigcirc$ (``next"), $\mathsf{U}$ (``until"), $\square$ (``always"), and $\diamondsuit$ (``eventually"). LTL formulas are interpreted over infinite words, as those generated by the system. For example, the word $w(0),w(1),\ldots$ where $w(0) = \{\pi_1, \pi_2\}, w(1) = \{\pi_1, \pi_2,\pi_3\}$, and $w(2),w(3),\ldots=\{\pi_1, \pi_4\}$ satisfies formulas $\square \pi_1, \diamondsuit \pi_3, \diamondsuit \square (\pi_1 \wedge \pi_4)$, and $\pi_2 \mathsf{U} \pi_4$ but violates $\square \pi_2$ and $\diamondsuit \pi_5$. We say that a trajectory $x(0),x(1),\ldots$ satisfies an LTL formula $\phi$ if and only if the corresponding word $w(0),w(1),\ldots$ satisfies $\phi$. The device satisfies $\phi$ from a given region $X \subseteq \mathcal{X}$ if and only if all trajectories originating in $X$ satisfy the formula.

We are now ready to formulate the main problem we consider in this paper:
\begin{problem}\label{problem:main}
Given a device constructed from characterized parts and a specification expressed as an LTL formula over a set of linear inequalities in the concentrations of proteins, determine if the device satisfies the specification.
\end{problem}

Our approach to Problem \ref{problem:main} consists of two main steps. Given a device, we first use the characterization data available for its parts to automatically construct a mathematical model by applying the procedure we develop in Sec. \ref{sec:identification}. The particular model structure we enforce allows us to guarantee that all experimental observations can be reproduced by the identified model. As a second step, we also exploit this structure to analyze the model from the temporal logic specification using a method inspired by model-checking, which we described in \cite{Yordanov2010} but review and extend in Sec. \ref{sec:analysis}. Our analysis procedure results in the computation of a satisfying (respectively, violating) region - a subset of the system's state space from which all trajectories are guaranteed to satisfy (respectively, violate) the specification. A device design is considered ``good" if analysis reveals a large satisfying region and an empty or small violating region, while a design is ``bad" whenever a substantial violating region is found. Given a library of characterized parts, our overall procedure can serve to evaluate a large number of possible device designs in order to select few for further experimental testing.

\section{MODEL CONSTRUCTION}\label{sec:identification}
In this section, we describe our procedure for the automatic construction of device models from part characterization data. As it will become clear later, the resulting models capture all experimental observations and take the form of uncertain parameter systems with different dynamics in different regions of the state space.

In Sec. \ref{sec:problem} we considered a simplified mechanism of gene expression (Fig. \ref{fig:pic1}). A gene $g$ was expressed from promoter $p$ at rate $\beta_p$ to make protein whose concentration was denoted by $x_g$ and which degraded at rate $\alpha_g$. We can express the dynamics of protein concentration as
\begin{equation}\label{eqn:const_prom}
x_g(k+1) = \alpha_gx_g(k) + \beta_p.
\end{equation}
In the problem formulation of Sec. \ref{sec:problem}, we assumed that, for each gene (protein) $g$,  $\alpha_g$ has a fixed value which is known for characterized parts, but $\beta_p$ is allowed to vary in some range, which is unknown and must be computed from the promoter characterization data (Fig. \ref{fig:formulation_pics}).

We first consider the computation of a range $B^c_{p}\subset \mathbb{R}$ for a constitutive promoter $p$, such that $\beta_{p} \in B^c_{p}$ in Eqn. (\ref{eqn:const_prom}). Then, we consider a regulated promoter where $\beta_{p} \in B_{p}(x_{g'})$  ({\em i.e.} the range of allowed rates $B_{p}(x_{g'})\subset \mathbb{R}$ is, in general, a function of the regulator concentration $x_{g'}$). For both, we first discuss the case when experimental observations of expression rates are directly available and later extend our procedure to compute such measurements indirectly from more realistic experimental data. We conclude the section by discussing the construction of models for general devices composed of a number of characterized parts. In developing our model identification procedure, we seek to compute a range of expression rates that is tight but contains all experimental measurements. This leads to the construction of models that can reproduce all observed behavior but are not overly general, which would make their subsequent analysis in Sec. \ref{sec:analysis} too conservative. In the following, we denote measured expression rates and protein concentrations from promoter $p$ and gene $g$ by $\tilde{\beta}_p$ and $\tilde{x}_g$, respectively.

\subsection{Constitutive promoter}\label{sec:const}
If promoter $p$ is constitutive, expression rate $\beta_p$ does not depend on the concentrations of other species in the system (there are no regulators) but varies in range $B^c_{p}$. If a data set $D^c_p = \{\tilde{\beta}_p(1),\ldots \tilde{\beta}_p(n)\}$ of experimentally measured expression rates is available (Fig. \ref{fig:pic2}), this range is simply
\begin{equation}\label{eqn:range}
B^c_p = [min(D^c_p),max(D^c_p)].
\end{equation}
This captures all experimentally observed rates and extrapolates under the assumption that any rate between the minimal and maximal observed one is also possible for the system.

In general, expression rates cannot be measured directly and must be computed from protein concentration measurements~\cite{Rosenfeld2005}. If gene $g$ is expressed from constitutive promoter $p$, given a finite trajectory fragment $\tilde{x}_g(0),\tilde{x}_g(1),\ldots,\tilde{x}_g(n+1)$ observed in experiments, from Eqn. (\ref{eqn:const_prom}) it follows that the expression rate $\tilde{\beta}_p(k)$ observed at step $k=0,\ldots,n$ is
\begin{equation}\label{eqn:const_data}
\tilde{\beta}_p(k)  = \tilde{x}_g(k+1) - \alpha_g\tilde{x}_g(k).
\end{equation}

The computation outlined above works when measurements of protein concentrations $\tilde{x}_g$ from individual cells are available. In experimental settings, it is often convenient to use techniques where protein concentrations for a large number of cells are measured simultaneously. In this case, individual cells are not identified uniquely and we must allow the possibility of a cell making a transition from the lowest observed concentration at a step $k$ to the highest one observed at step $k+1$ and vice versa. Then, a minimal and a maximal possible rate is computed at each step and included in the set $D^c_p$ but the rest of the computation remains the same. Single cell experimental techniques lead to the identification of tighter expression rate ranges and we only consider such measurements through the rest of this paper.

\subsection{Regulated promoter}\label{sec:reg}
For a regulated promoter $p$, the rate of expression $\beta_p$ varies in a range $B_p(x_{g'})$, which is a function of the regulator concentration $x_{g'}$. Range $B_p(x_{g'})$ is unknown and must be computed from the available promoter characterization data $(\tilde{\beta}_p,\tilde{x}_{g'}) \in D_p$ ({\em i.e.} $D_p$ is a set of expression rates measured at different repressor concentrations as in Fig. \ref{fig:pic3}). In the following, we focus on the construction of the set
\begin{equation}
\bar{B}_p = \{(\beta_p,x_{g'})\;|\;x_{g'}\in [x_{g'}^{min},x_{g'}^{max}], \beta_p \in B_p(x_{g'})\}.
\end{equation}
This allows us to compute $B_p(x_{g'})$ at arbitrary concentrations $x_{g'}$ as the slice of $\bar{B}_p$ at $x_{g'}$ ({\em i.e.} $B_p(x_{g'}) = \{\beta_p\;|\;(\beta_p,x_{g'}) \in \bar{B}_p\}$). By constructing the tightest $\bar{B}_p$ that contains all experimental measurements ({\em i.e.} $D_p \subset \bar{B}_p$), we guarantee that the model we identify can reproduce all observed behavior but our subsequent analysis in Sec. \ref{sec:analysis} is not overly conservative.

\begin{figure*}[ht]
\centering
\subfigure[Constant range]{\includegraphics[scale=0.33]{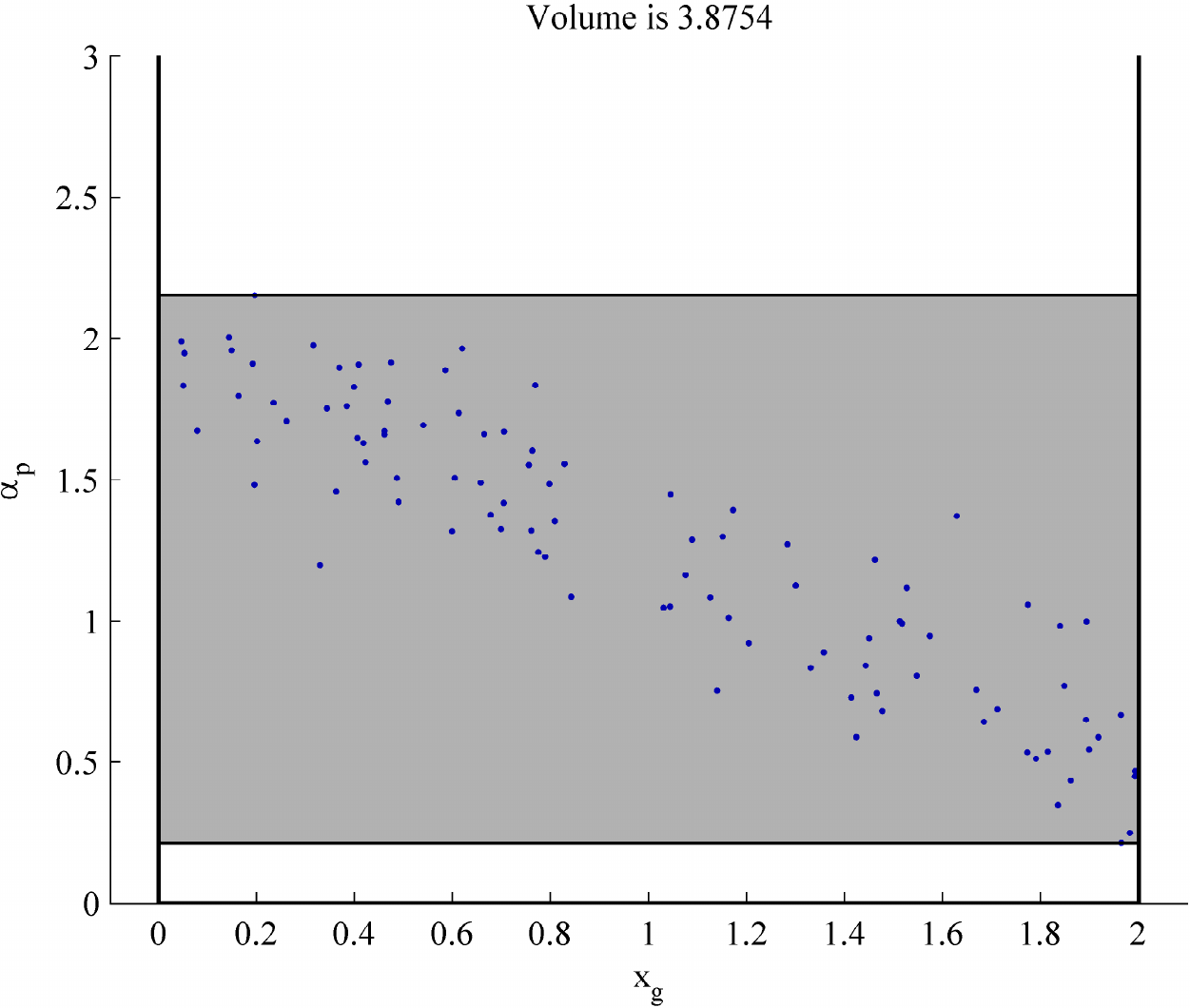}\label{fig:fit0}}
\subfigure[Piecewise constant range]{\includegraphics[scale=0.33]{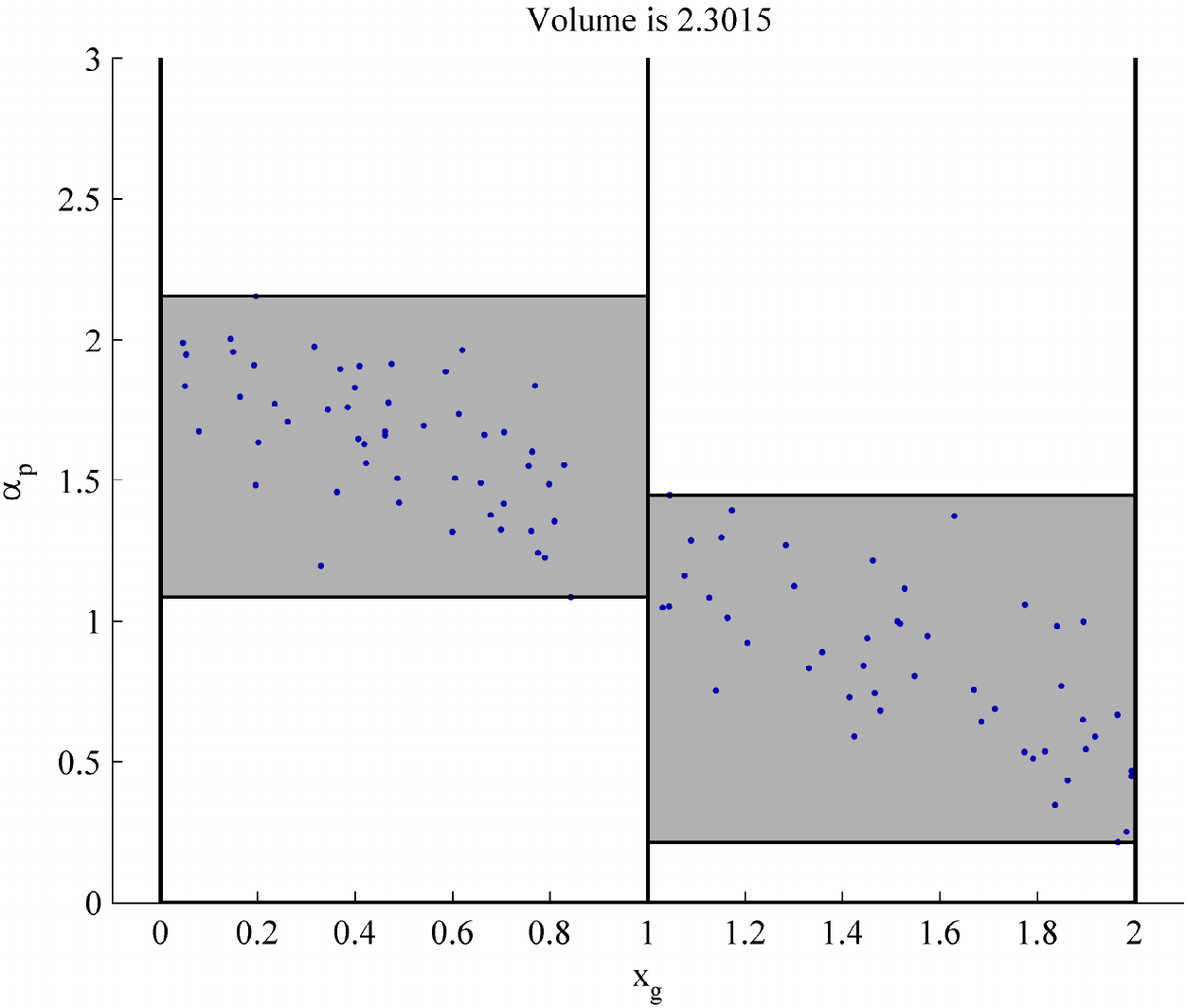}\label{fig:fit1}}
\subfigure[Piecewise linear range]{\includegraphics[scale=0.33]{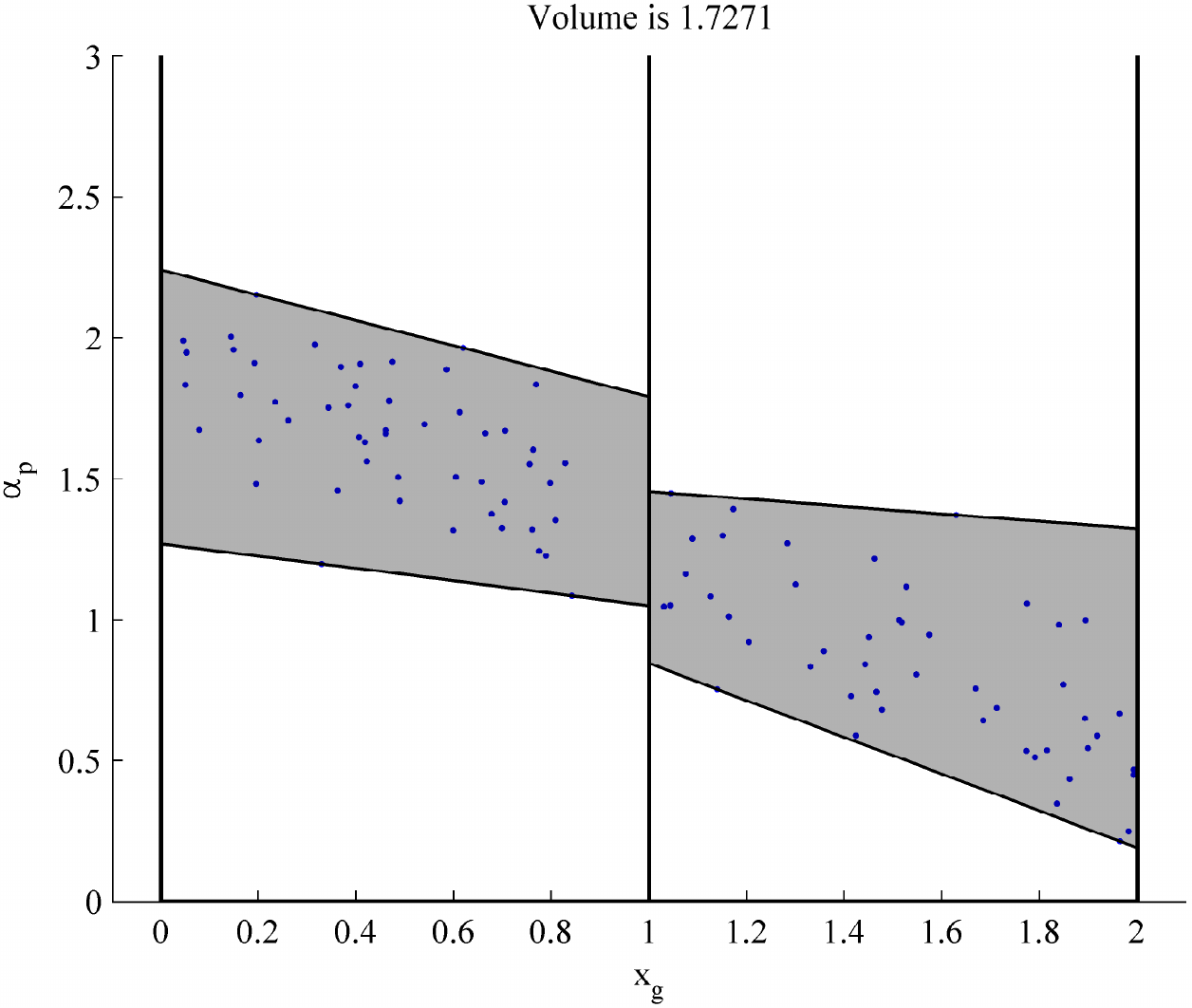}\label{fig:fit2}}
\subfigure[Convex hull range]{\includegraphics[scale=0.33]{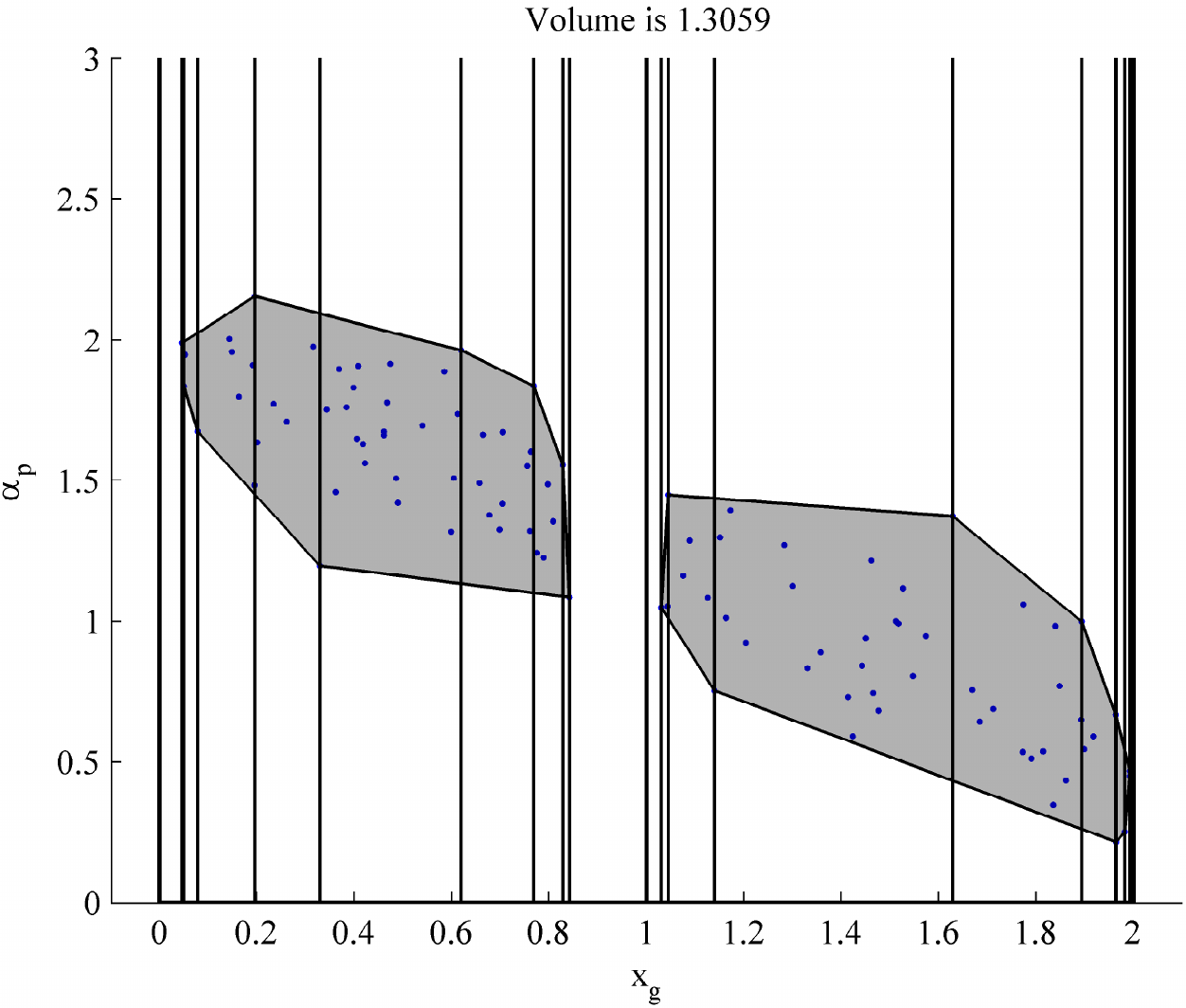}\label{fig:fit3}}
\caption{Different methods for fitting experimentally observed expression rates (the same data is used for all procedures). Initial thresholds are shown as thick vertical lines. Additional thresholds for \subref{fig:fit3} are shown as thin vertical lines. Resulting regions from each procedures are shown in gray.}\label{fig:fit_pics}
\vspace{-0.25in}
\end{figure*}

It is most straightforward to extend Eqn. (\ref{eqn:range}) and compute a {\em constant range}  (Fig. \ref{fig:fit0}) as
\begin{eqnarray}
\bar{B}_p = [min(\hat{D}_p),max(\hat{D}_p)] \times  [x_{g'}^{min},x_{g'}^{max}] \mbox{ where} \label{eqn:regulated_range} \\
\hat{D}_p = \{\tilde{\beta}_p\;|\;\exists x_{g'}\in  [x_{g'}^{min},x_{g'}^{max}], (\tilde{\beta}_p,\tilde{x}_{g'}) \in D_p\} \label{eqn:regulated_data}
\end{eqnarray}
In this case, the range $B_p(x_{g'})$  is the same for all regulator concentrations $x_{g'}$ and, although it captures all experimental data, it includes expression rates from both an activated and repressed $p$ which makes the overall method conservative.

To compute a tighter range, we introduce a set of thresholds $\theta^i_{g'}$ such that $x_{g'}^{min} \leq \theta^i_{g'} \leq x_{g'}^{max}$ for all $i=1,\ldots,n_{g'}$ and $\theta^i_{g'} < \theta^{i+1}_{g'}$ for all $i=1,\ldots,n_{g'}-1$. We discuss the computation of these thresholds in Sec. \ref{sec:multi} and, in this subsection, we focus on the expression rates observed when regulator concentration falls in the region between two thresholds (see Fig. \ref{fig:fit_pics}). For $i=1,\ldots,n_{g'}-1$, we define the subset
\begin{equation}
D^i_p =\{(\tilde{\beta}_p,\tilde{x}_{g'}) \in D_p\;|\; \theta^i_{g'}\leq \tilde{x}_{g'}\leq \theta^{i+1}_{g'}\}. \nonumber
\end{equation}
By applying Eqn. (\ref{eqn:regulated_range}) locally in each region, we can compute a {\em piecewise constant range} (Fig. \ref{fig:fit1}) as
\begin{eqnarray}
\bar{B}_p   & = & \bigcup^{n_{g'}-1}_{i=1}\bar{B}^i_p \mbox{ where } \nonumber \\
\bar{B}^i_p & = & [min(\hat{D}^i_p),max(\hat{D}^i_p)] \times [\theta^i_{g'},\theta^{i+1}_{g'}] \nonumber
\end{eqnarray}
and $\hat{D}^i_p$ is computed for each subset $D^i_p$ as in Eqn. (\ref{eqn:regulated_data}).

A piecewise constant range captures all experimental observations while allowing different local ranges for different regulator concentration regions. This procedure also leads to some computational advantages in Sec. \ref{sec:analysis} but it might still be too conservative ({\em i.e.} the volume of $\bar{B}_p$ might be too large). We can also compute 
\begin{equation}\label{eqn:hull_range}
\bar{B}^i_p = hull(D^i_p)
\end{equation}
which is the smallest convex set containing all observed expression rates in each region (shown in Fig. \ref{fig:fit3}) and, therefore, $\bar{B}_p$ has minimal volume. As it will become clear later, for such {\em convex hull range} $B_p(x_{g'})$ cannot be computed easily unless additional thresholds at each vertex are introduced, which leads to complications.

As a compromise, a {\em piecewise linear range} as in Fig. \ref{fig:fit2} has, in general, smaller volume than the piecewise constant range from Fig. \ref{fig:fit1} and does not require additional thresholds as the convex hull range from Fig. \ref{fig:fit3}. Such a range can be computed by enumerating all trapezoids that have the two thresholds and two of the supporting planes from the convex hull range as sides, and selecting the one with the smallest volume. Under the additional assumption that expression rates are measured only at regulator concentrations that fall on thresholds, the procedures from Figs. \ref{fig:fit2} and \ref{fig:fit3} result in the same set $\bar{B}_p$, where each $\bar{B}_p^i$ can be computed using Eqn. (\ref{eqn:hull_range}).

For a piecewise linear range (Fig. \ref{fig:fit2}), given regulator concentration $x_{g'}$ such that $x_{g'} = \lambda \theta^i_{g'} + (1-\lambda)\theta^{i+1}_{g'}$ for some $i=1,\ldots,n_{g'}-1$ and $\lambda \in [0,1]$, we have
\begin{equation}\label{eqn:range_comput_linear}
B_p(x_{g'}) = \lambda B_p(\theta^i_{g'}) + (1-\lambda)B_p(\theta^{i+1}_{g'}).
\end{equation}

As for constitutive promoters, when expression rates are not available directly, they can be computed from protein concentration measurements. Given genes $g$,$g'$ and a promoter $p$, such that $g$ is expressed from $p$ and $g'$ regulates $p$, and a trajectory fragment $\tilde{x}(0),\tilde{x}(1),\ldots,\tilde{x}(n+1)$ where $\tilde{x}(k) = (\tilde{x}_{g}(k), \tilde{x}_{g'}(k))$ is a vector of regulator and protein concentrations, we have
\begin{eqnarray}\label{eqn:reg_data}
D_p = \{(\tilde{\beta}_p(k),\tilde{x}_{g'}(k))\;|\;\tilde{x}(k) = (\tilde{x}_{g}(k),\tilde{x}_{g'}(k)), \\
      \tilde{\beta}_p(k) = \tilde{x}_{g}(k+1) - \alpha_{g}\tilde{x}_{g}(k), k=1,\ldots,n\}. \nonumber
\end{eqnarray}

\subsection{Device models}\label{sec:multi}
To summarize the construction of models using the procedures we discussed so far, we consider a device consisting of a set of genes $G$ and promoters $P$ (see the problem formulation in Sec. \ref{sec:problem}). For notational simplicity, we assume that for $i=1,\ldots,N$, gene $g_i \in G$ is expressed from promoter $p_i \in P$, which is either constitutive or regulated by the protein produced by gene $g'_i \in G$. We assume that, for each gene $g\in G$, we have at least two thresholds ({\em i.e.} $n_g\geq 2$) where $\theta^1_g = x_g^{min}$ and $\theta^{n_g}_g=x_g^{max}$ ({\em i.e.} the boundaries of the state space $\mathcal{X}$ introduced in Sec. \ref{sec:problem} are thresholds). Computing the set of thresholds is not the focus of this paper but related methods are available \cite{Drulhe08}. Here, we implement a sampling procedure where, out of a number of randomly generated thresholds, we select the subset of a given size that minimizes the volume of $\bar{B}_p$.

For a state  $x \in \mathcal{X}_l$ where $x = (x_{g_1},\ldots,x_{g_N})$, the dynamics of each component $x_g$ are given by Eqn. (\ref{eqn:const_prom}) where
\begin{equation}
\beta_{p_i} \in \left \{ \begin{array}{l l}
B^c_p & \mbox{if $p$ is constitutive or } \\
B_p(x_{g'_i}) & \mbox{if $p$ is regulated}
\end{array} \right.
\end{equation}

It is important to note that the identified model can reproduce all experimental data used for part characterization. Consider a trajectory fragment used in Eqn. (\ref{eqn:const_data}) or (\ref{eqn:reg_data}) to respectively characterize a constitutive or regulated promoter. We can guarantee that the expression rate from the promoter, required to reach the concentration of the expressed protein observed at step $k+1$ starting from the concentration observed at step $k$, is always in the allowed range. In Sec. \ref{sec:analysis} we will show that the model structure is different for piecewise constant and piecewise linear ranges but allows the computation of finite abstractions through polyhedral operations, enabling the application of formal analysis techniques.

\begin{remark}\label{remark:unique}
Note that range $B_p(\theta^i_{g'})$ is not well defined and might be different for the regions that share threshold $\theta^i_{g'}$ (for example, see Figs. \ref{fig:fit1} and \ref{fig:fit2}) and, thorough the rest of this paper, we only consider states from the interior of regions.
\end{remark}

\section{FORMAL ANALYSIS}\label{sec:analysis}
In Sec. \ref{sec:identification} we developed a procedure for the automatic construction of device models from part characterization data. All experimental measurements were captured in the resulting models by allowing expression rates to vary in certain ranges. In this section we show that, despite this uncertainty, finite quotients of the identified models can be constructed using polyhedral operations, which enables analysis through methods inspired by model checking. With the exception of Prop. \ref{prop:post_comput}, the material presented in this section is largely a review of our results from \cite{Yordanov2010}.

The state space $\mathcal{X}$ from Eqn. (\ref{eqn:state_space}) is partitioned by the thresholds $\theta^i_g, i=1,\ldots,n_g$ of all genes $g \in G$ into a number of hyper-rectangular regions. We partition $\mathcal{X}$ further using all linear inequalities $\pi \in \Pi$ (Eqn. (\ref{eqn:ineq})) and ignore the measure-zero set consisting of all boundaries.\footnote{It is unreasonable to assume that equality constraints can be detected in practice and, in general, trajectories of the system do not start from or disappear in this measure-zero set.} This results in a set of open polytopes $\mathcal{X}_l, l \in L$ such that, for all $l_1, l_2 \in L$, $\mathcal{X}_{l_1} \cap \mathcal{X}_{l_2} = \emptyset$ and $\cup_{l\in L}cl(\mathcal{X}_l) = \mathcal{X}$, where $cl()$ denotes the closure of a set. We denote the set $\cup_{l\in L}\mathcal{X}_l$ as $\bar{\mathcal{X}}$. Note that all states from a given region satisfy the same atomic propositions ({\em i.e.} for all $x_1,x_2 \in \mathcal{X}_l$ for some $l \in L$ and all $\pi \in \Pi$, $x_1\vDash \pi$ if and only if $x_2\vDash \pi$).

We define two states as equivalent if and only if they belong to the same region $\mathcal{X}_l$ for some $l \in L$. The finite, proposition preserving quotient induced by this equivalence relation is the transition system $T=(Q,\rightarrow,\Pi,\vDash)$ where
\begin{itemize}
 \item $Q=L$ is the finite set of states,
 \item $\rightarrow \subseteq Q \times Q$ is the transition relation defined as $(l_1,l_2) \in \rightarrow$ if and only if there exists a transition from a state in region $\mathcal{X}_{l_1}$ to a state in $\mathcal{X}_{l_2}$,
 \item $\Pi$ is the set of atomic propositions from Eqn. (\ref{eqn:ineq}), and
 \item $\vDash \subseteq Q\times \Pi$ is the satisfaction relation\footnote{We abuse the notation and use symbol $\vDash$ to denote the satisfaction of a proposition by a state in the original infinite system and its abstraction $T$.} where, given $l \in L$ and $\pi \in \Pi$, $l \vDash \pi$ if and only if, for all $x \in \mathcal{X}_l$, $x \vDash \pi$.
\end{itemize}
From the definition of the transition relation $\rightarrow$, it follows that $T$ simulates the infinite system identified through our procedure from Sec. \ref{sec:identification} (in other words, $T$ can produce any word that the infinite system can produce \cite{Milner1989}). This allows us to guarantee that if an arbitrary LTL formula $\phi$ is satisfied by $T$ at state $l\in L$, then all trajectories of the system originating in region $\mathcal{X}_l$ satisfy the formula. Note that when $T$ does not satisfy $\phi$ from state $l$ we cannot say anything about the satisfaction of $\phi$ from region $\mathcal{X}_l$, which makes the overall method conservative.

In \cite{Yordanov2010} we developed an analysis procedure based on the construction, model checking and refinement of simulation quotients such as $T$. Our algorithm used model checking to partition the set of states $L$ into set $L^\phi \subseteq L$ from which $T$ satisfied an LTL formula $\phi$ and $L^{\neg\phi} \subseteq L$ from which $T$ satisfied the negation $\neg\phi$. This allowed us to guarantee that all trajectories originating in the {\em satisfying region} $\mathcal{X}^\phi =\bigcup_{l \in L^\phi}\mathcal{X}_l$ and none of the trajectories originating in the {\em violating region} $\mathcal{X}^{\neg\phi} =\bigcup_{l \in L^{\neg\phi}}\mathcal{X}_l$ satisfied $\phi$. Both satisfying and violating trajectories originated in region
$\bar{\mathcal{X}}\setminus (\mathcal{X}^{\phi} \cup \mathcal{X}^{\neg\phi}$) and our algorithm implemented an iterative  refinement procedure to try and separate them, in which case $\mathcal{X}^{\phi}$ and $\mathcal{X}^{\neg\phi}$ can be expanded.

To apply our method from \cite{Yordanov2010} (implemented as the software tool \texttt{FaPAS}) we need to be able to construct $T$, which reduces to the computation of its transitions $\rightarrow$. For all $l \in L$, we denote the set of states reachable from $\mathcal{X}_l$ in one step as $Post(\mathcal{X}_l)$. Transitions of $T$ can be computed as
\begin{equation}\label{eqn:trans_comput}
(l_1,l_2) \in \rightarrow \mbox{ if and only if } Post(\mathcal{X}_{l_1}) \cap \mathcal{X}_{l_2} \neq \emptyset.
\end{equation}

To show that $T$ can be constructed, we show that $Post(\mathcal{X}_{l_1})\cap \mathcal{X}_{l_2}$ is computable for all $l_1,l_2 \in L$. We use the notation introduced in Sec. \ref{sec:identification} where each promoter, gene and regulator is denoted by $p_i\in P$ and $g_i, g'_i \in G$, $i=1,\ldots,N$, respectively. Given a state $x \in \mathcal{X}_l$ for some $l \in L$ such that $x = (x_{g_1},\ldots,x_{g_N})$, the overall system dynamics are given by
\begin{equation}\label{eqn:system_multiaffine}
x(k+1) \in  Ax(k) + B(x(k)),
\end{equation}
where $A$ is the diagonal matrix of degradation rates $A = diag(\alpha_{g_1},\ldots,\alpha_{g_N})$ and
\begin{eqnarray} \label{eqn:overall_range}
B(x) & = & B_1(x_{g'_1}) \times \ldots \times B_N(x_{g'_N}) \mbox{ where } \\
B_i(x_{g'_i}) & = & \left \{ \begin{array}{l l}
B_{p_i}^c & \mbox{if $p_i$ is constitutive or } \\
B_{p_i}(x_{g'_i}) & \mbox{if $p_i$ is regulated}
\end{array} \right. \nonumber
\end{eqnarray}

If the piecewise constant procedure from Sec. \ref{sec:identification} is used, for all states $x_1,x_2 \in \mathcal{X}_l, l \in L$, we have $B(x_1)= B(x_2)= B_l$. Therefore, the dynamics from Eqn. (\ref{eqn:system_multiaffine}) reduce to
\begin{equation}\label{eqn:system_affine}
x(k+1) \in Ax(k) + B_l \mbox{ when } x(k) \in \mathcal{X}_l
\end{equation}
and $Post(\mathcal{X}_{l})=A\mathcal{X}_l+B_l$.
\begin{proposition}\label{prop:post_comput}
For the more general case when the piecewise linear procedure from Sec. \ref{sec:identification} is used, for all $l \in L$, $Post(\mathcal{X}_l)$ is convex and computable as\footnote{As mentioned in Remark \ref{remark:unique}, the set $B(v)$ might be different for different regions that share vertex $v \in \mathcal{V}(\mathcal{X}_l)$ but from the index $l \in L$ it is always clear which $B(v)$ is used for the computation.}
\begin{equation}
Post(\mathcal{X}_l) = hull(\{Av + B(v)\;|\;v \in \mathcal{V}(\mathcal{X}_l)\}). \nonumber
\end{equation}
\begin{proof}
See Appendix.
\end{proof}
\end{proposition}

Following from the results presented so far, regardless of which procedure from Sec. \ref{sec:identification} is used, the intersection $Post(\mathcal{X}_{l_1})\cap\mathcal{X}_{l_2}$ is convex and computable for all $l_1,l_2 \in L$. Then, transitions can be computed using Eqn. (\ref{eqn:trans_comput})  which completes the construction of $T$ and, therefore, the satisfying and violation regions of the system identified in Sec. \ref{sec:identification} can be computed. The relative volumes of those regions can be used to determine if a device satisfies the specification, which provides a solution to Problem \ref{problem:main}. The same approach can also be used to compare different designs constructed from a set of parts based on their satisfaction of a common specification. To illustrate such an application, in Sec. \ref{sec:case_study} we use our method to design a synthetic gene network.

\section{DESIGN OF GENE NETWORKS}\label{sec:case_study}

\begin{figure}[t]
\centering
\includegraphics[scale=0.65]{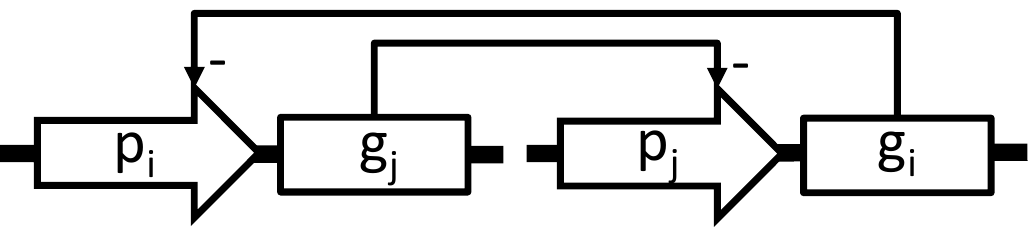}
\caption{Toggle switch. Two repressor proteins are expressed from two regulated promoters and mutually repress each other.}\label{fig:toggle}
\vspace{-0.25in}
\end{figure}

To illustrate our method, we apply it to the design of a bistable gene network inspired by the ``genetic toggle switch"~\cite{Gardner2000}, which has the topology shown in Fig. \ref{fig:toggle}. We start by constructing a library of parts, which includes three genes (denoted by $g_1,\ldots g_3$) and three promoters (denoted by $p_1,\ldots p_3$). We assume that the proteins produced from all three genes are stable and their degradation rates are negligible compared to the dilution due to cell growth, which leads to a protein half-life of $30$ min - an estimate of the generation time of bacteria. All promoters are regulated and the protein produced by gene $g_i$ represses promoter $p_i$.

\begin{figure*}[ht]
\centering
\subfigure{\includegraphics[scale=0.33]{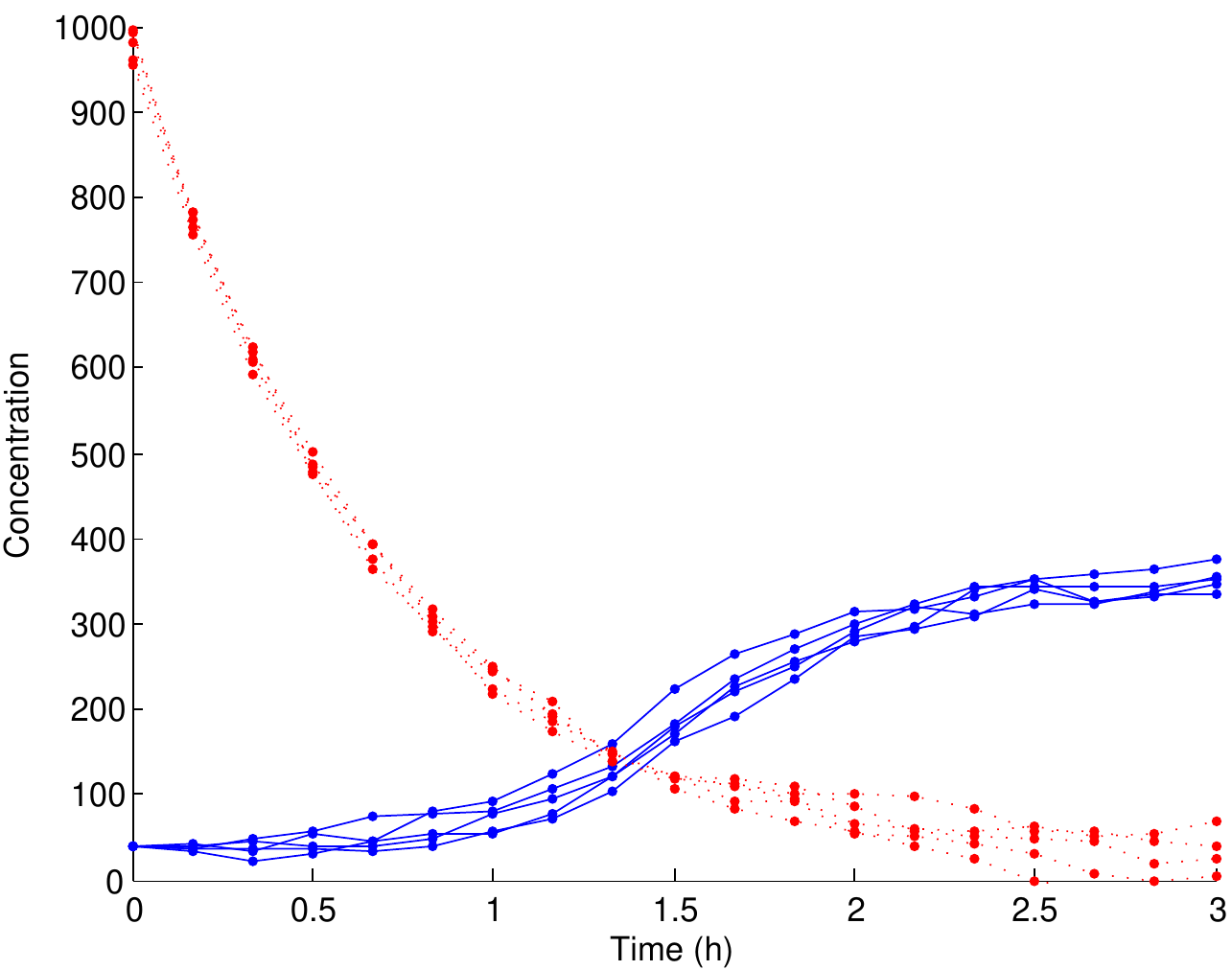}\label{fig:char_traj1}}
\subfigure{\includegraphics[scale=0.33]{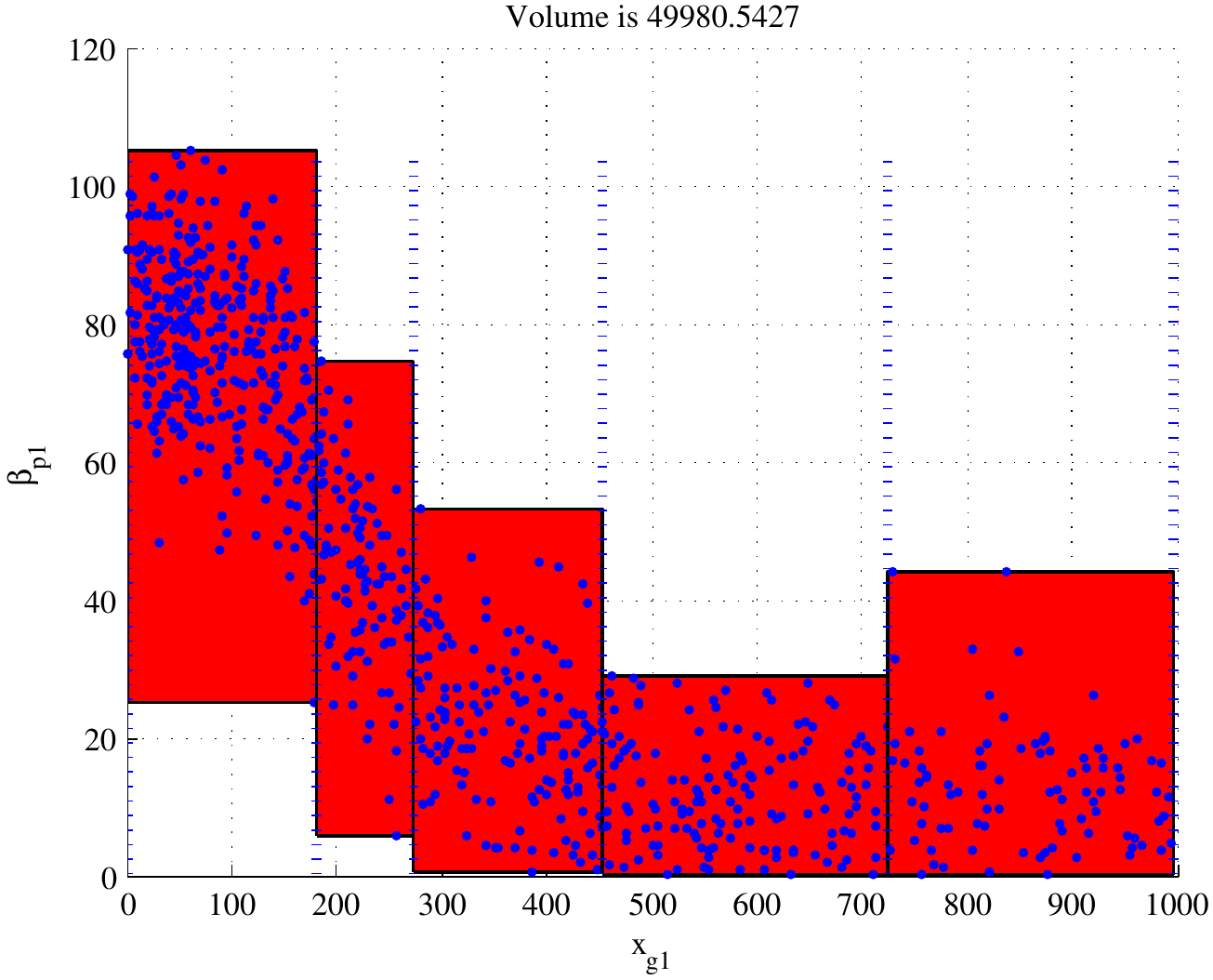}\label{fig:const1}}
\subfigure{\includegraphics[scale=0.33]{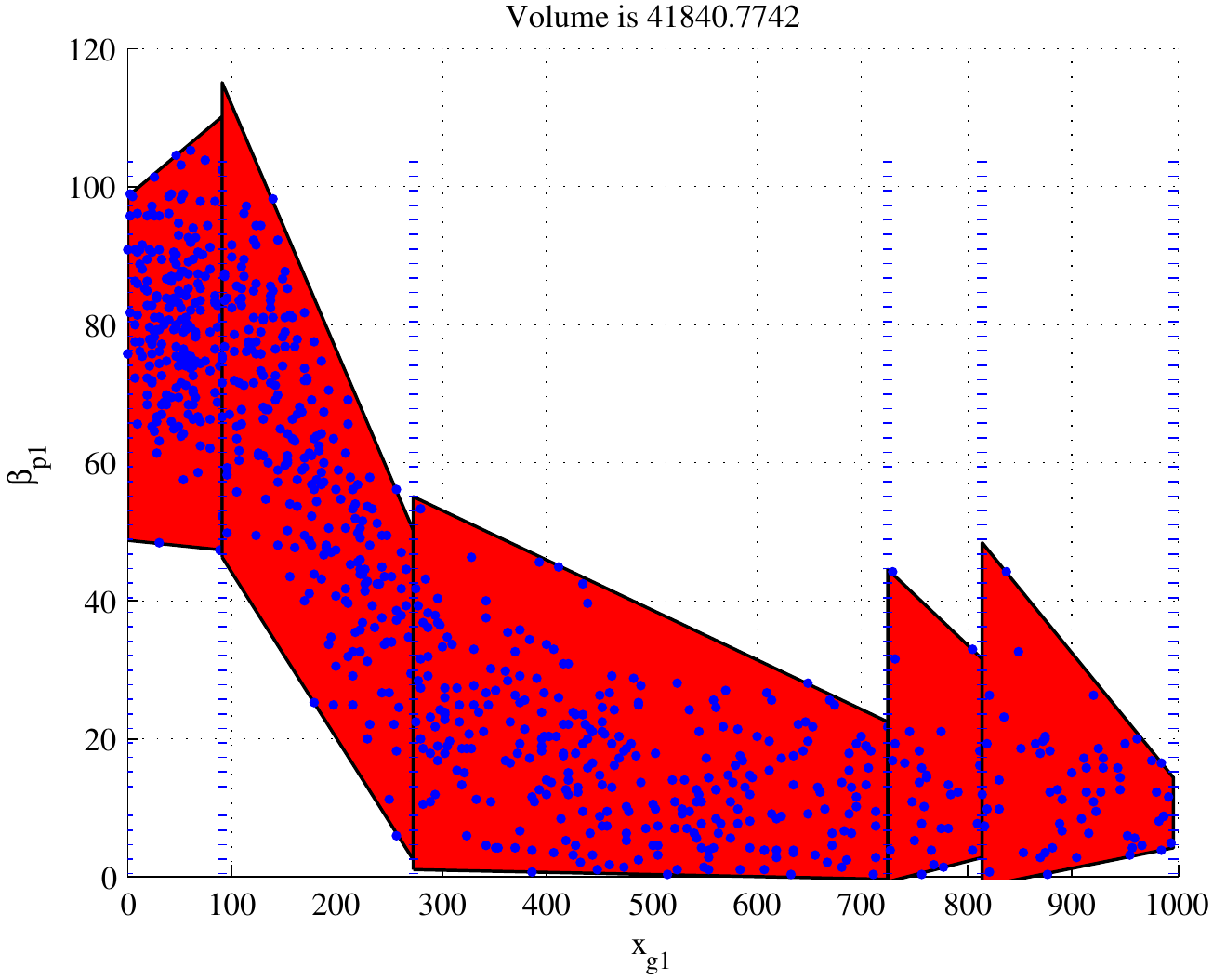}\label{fig:lin1}}
\subfigure{\includegraphics[scale=0.33]{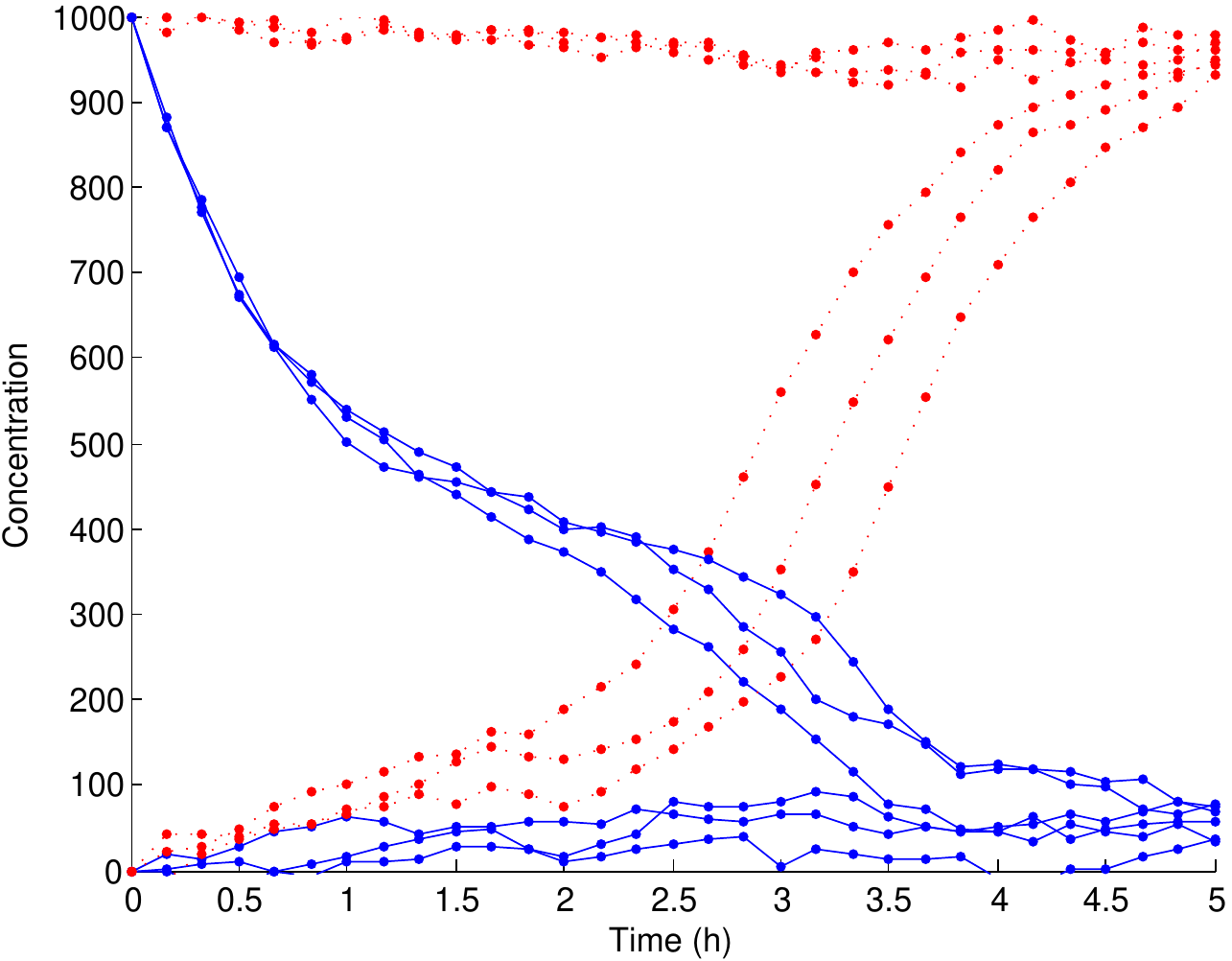}\label{fig:traj1}}

\subfigure{\includegraphics[scale=0.33]{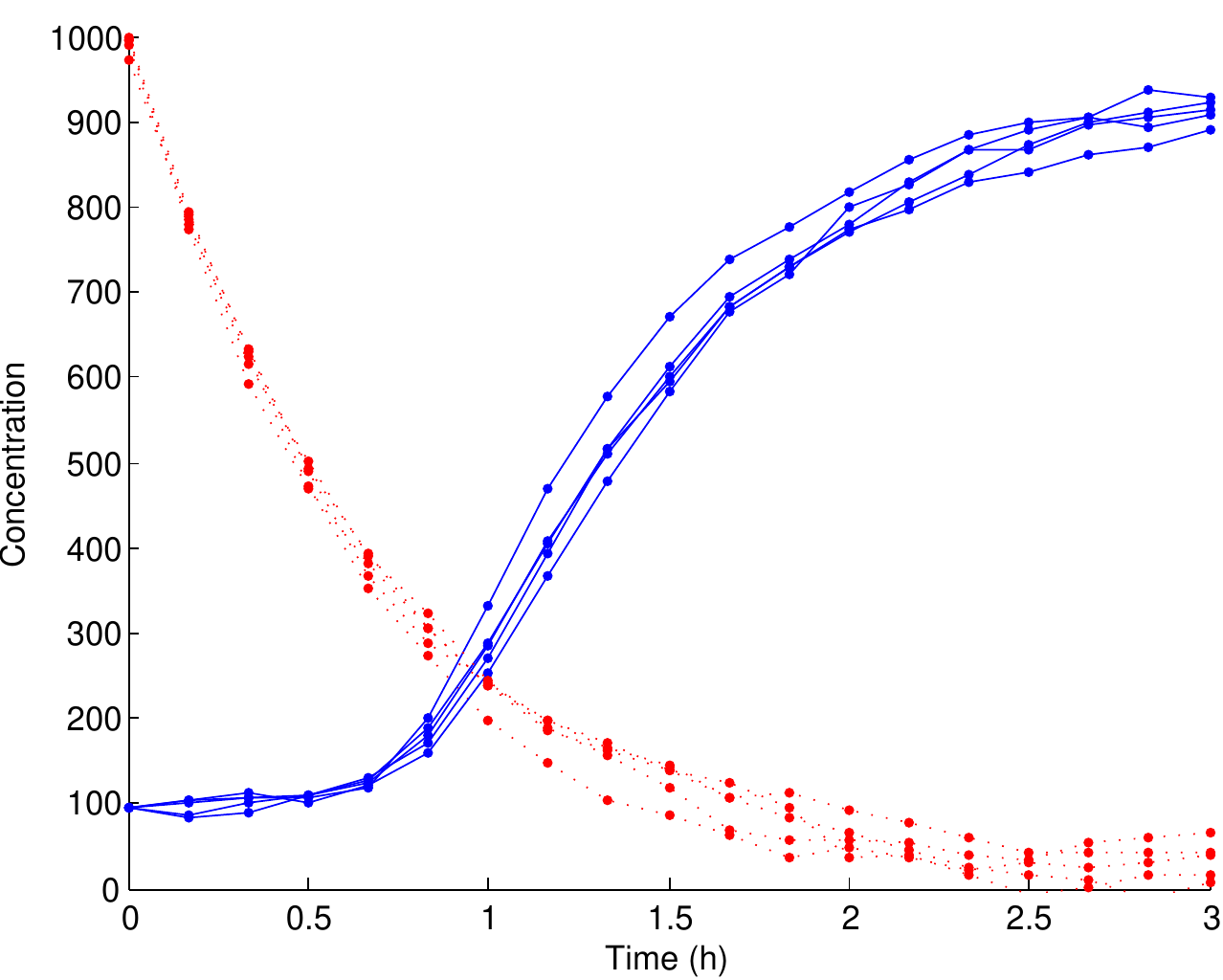}\label{fig:char_traj2}}
\subfigure{\includegraphics[scale=0.33]{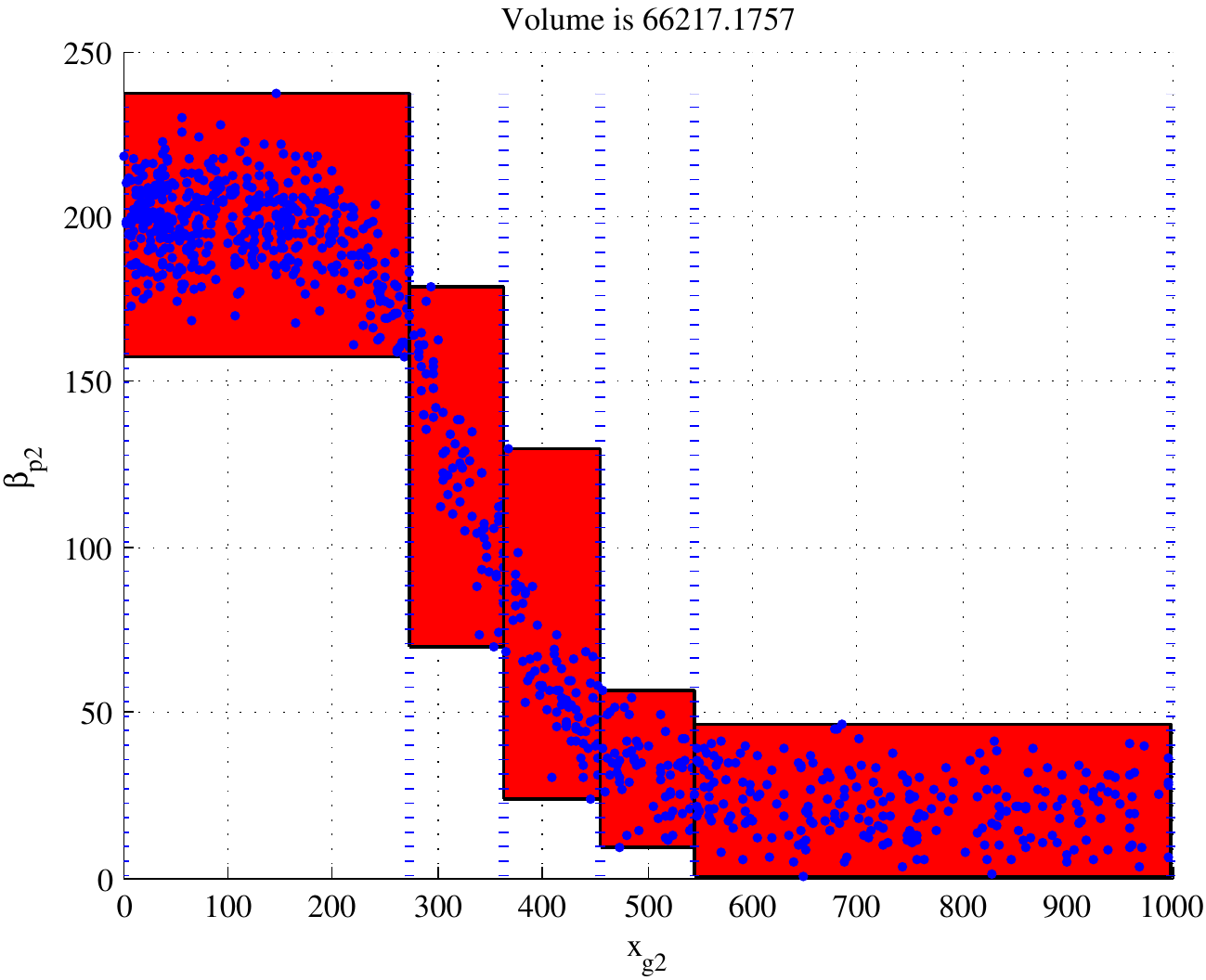}\label{fig:const2}}
\subfigure{\includegraphics[scale=0.33]{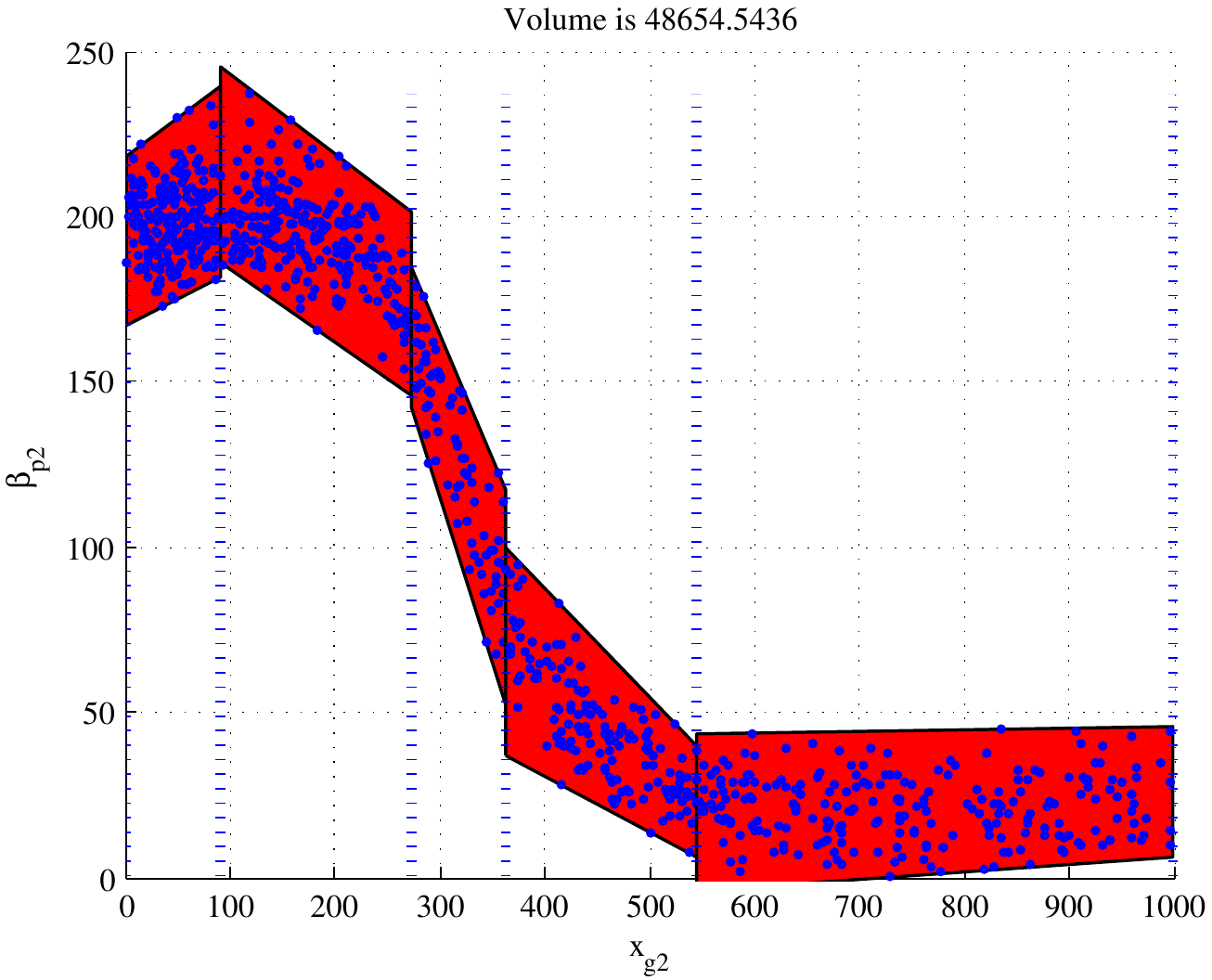}\label{fig:lin2}}
\subfigure{\includegraphics[scale=0.33]{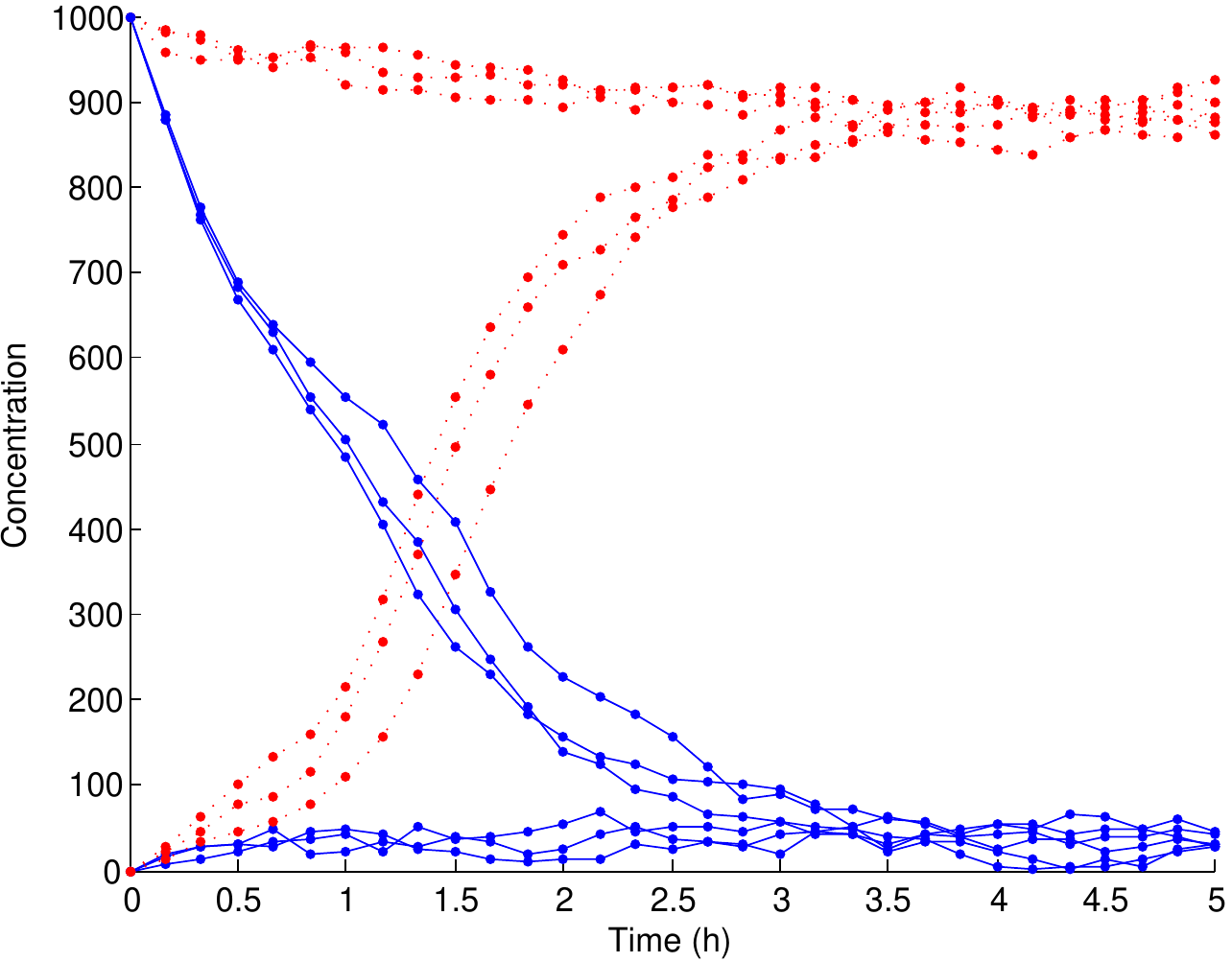}\label{fig:traj2}}

\subfigure{\includegraphics[scale=0.33]{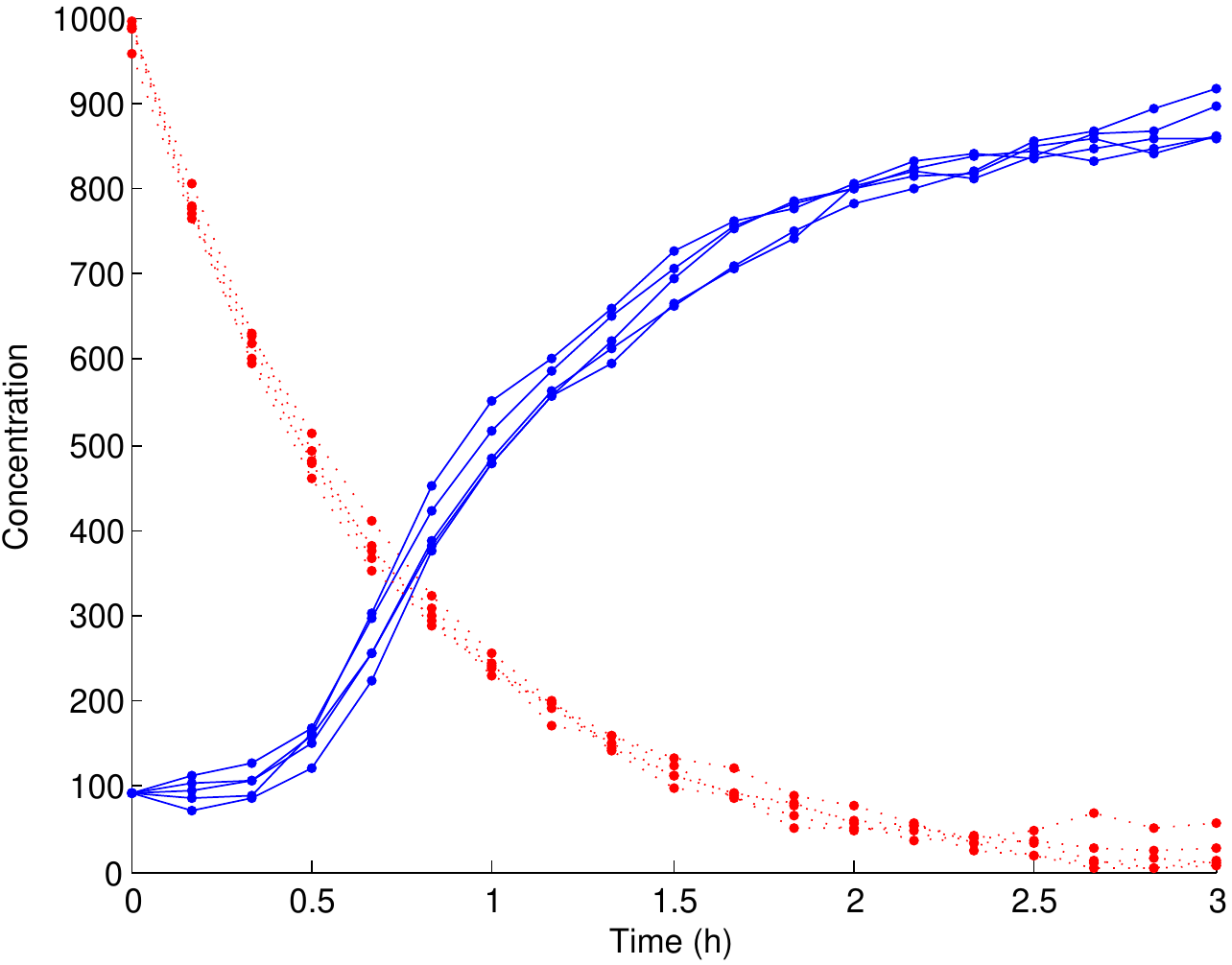}\label{fig:char_traj3}}
\subfigure{\includegraphics[scale=0.33]{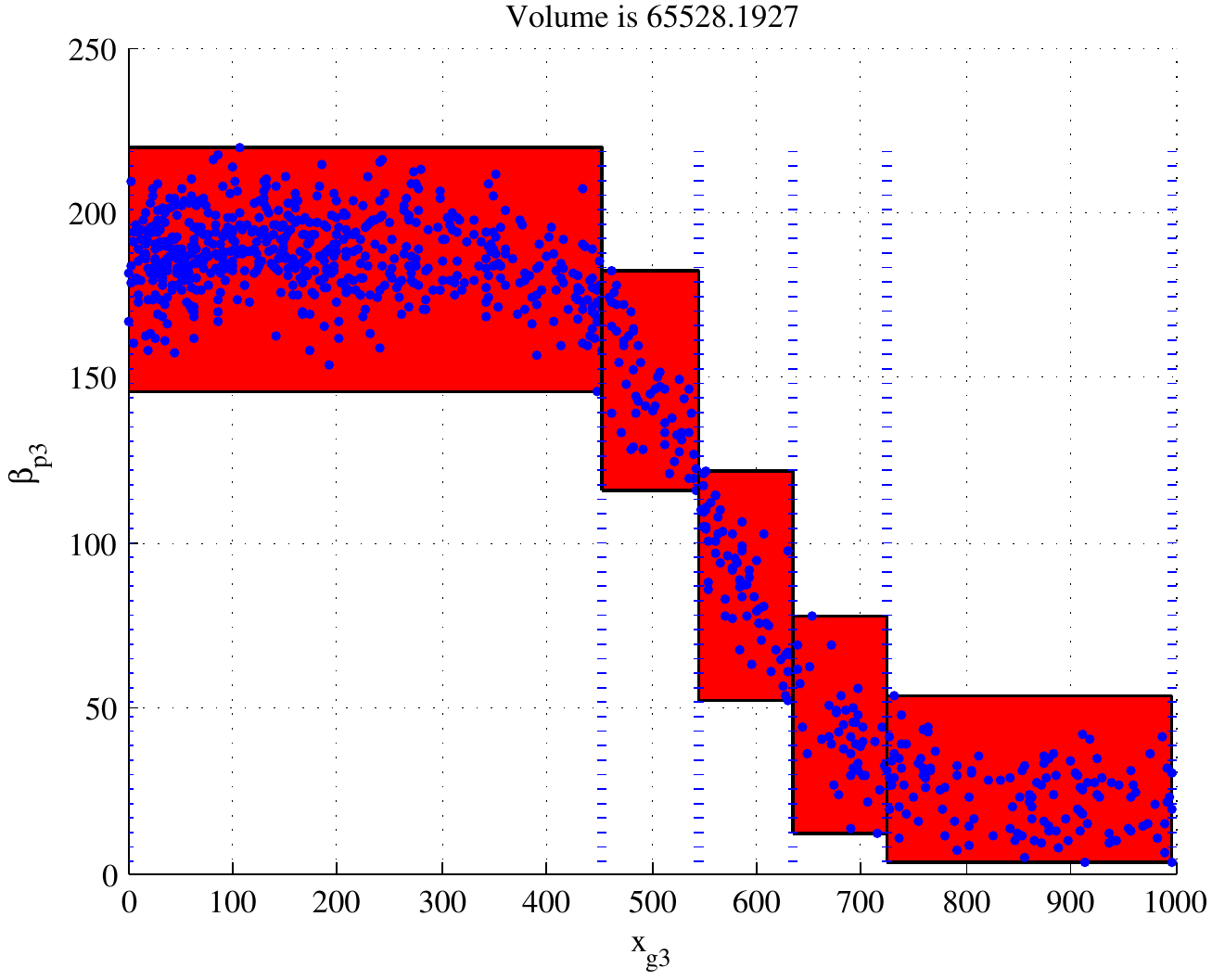}\label{fig:const3}}
\subfigure{\includegraphics[scale=0.33]{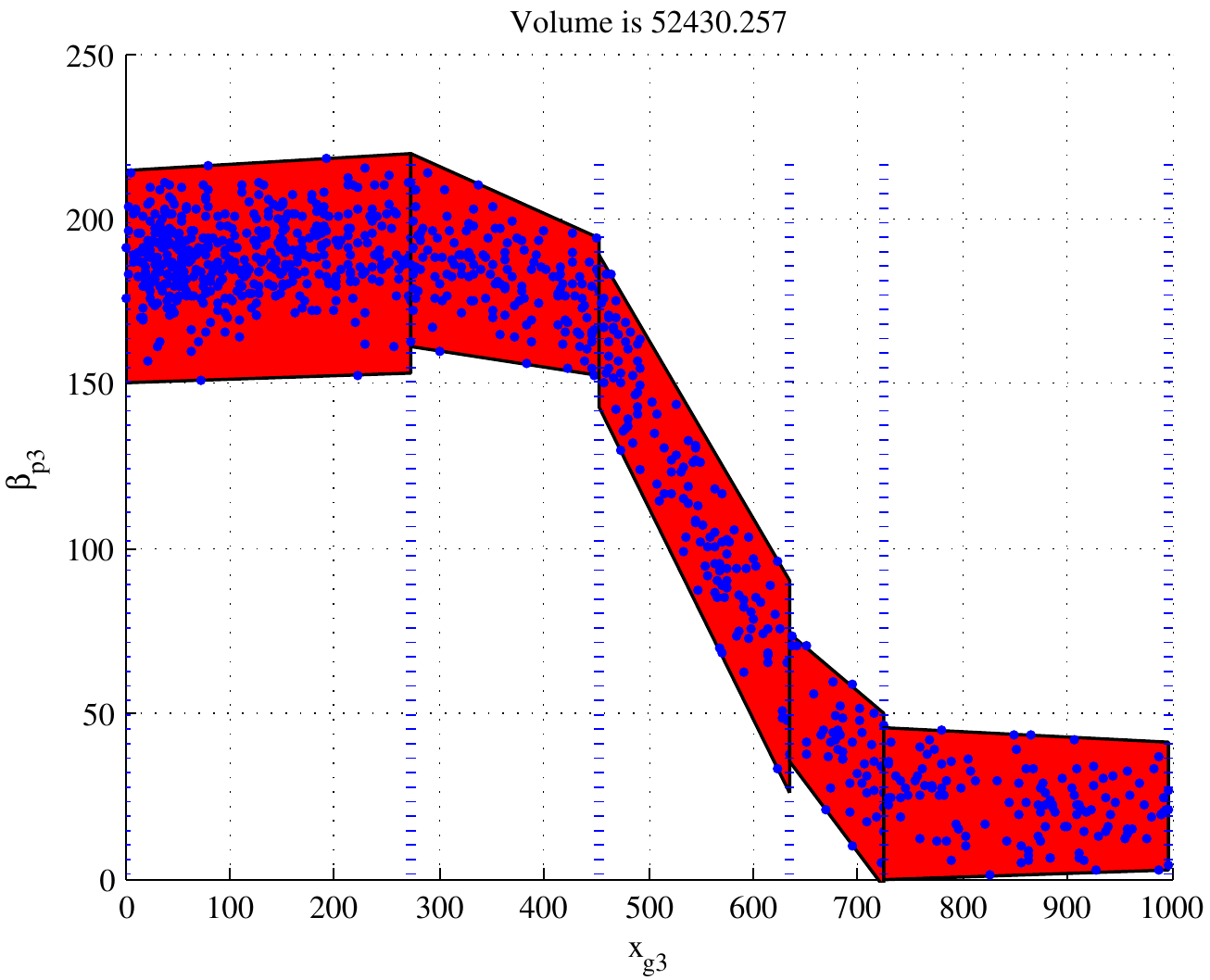}\label{fig:lin3}}
\subfigure{\includegraphics[scale=0.33]{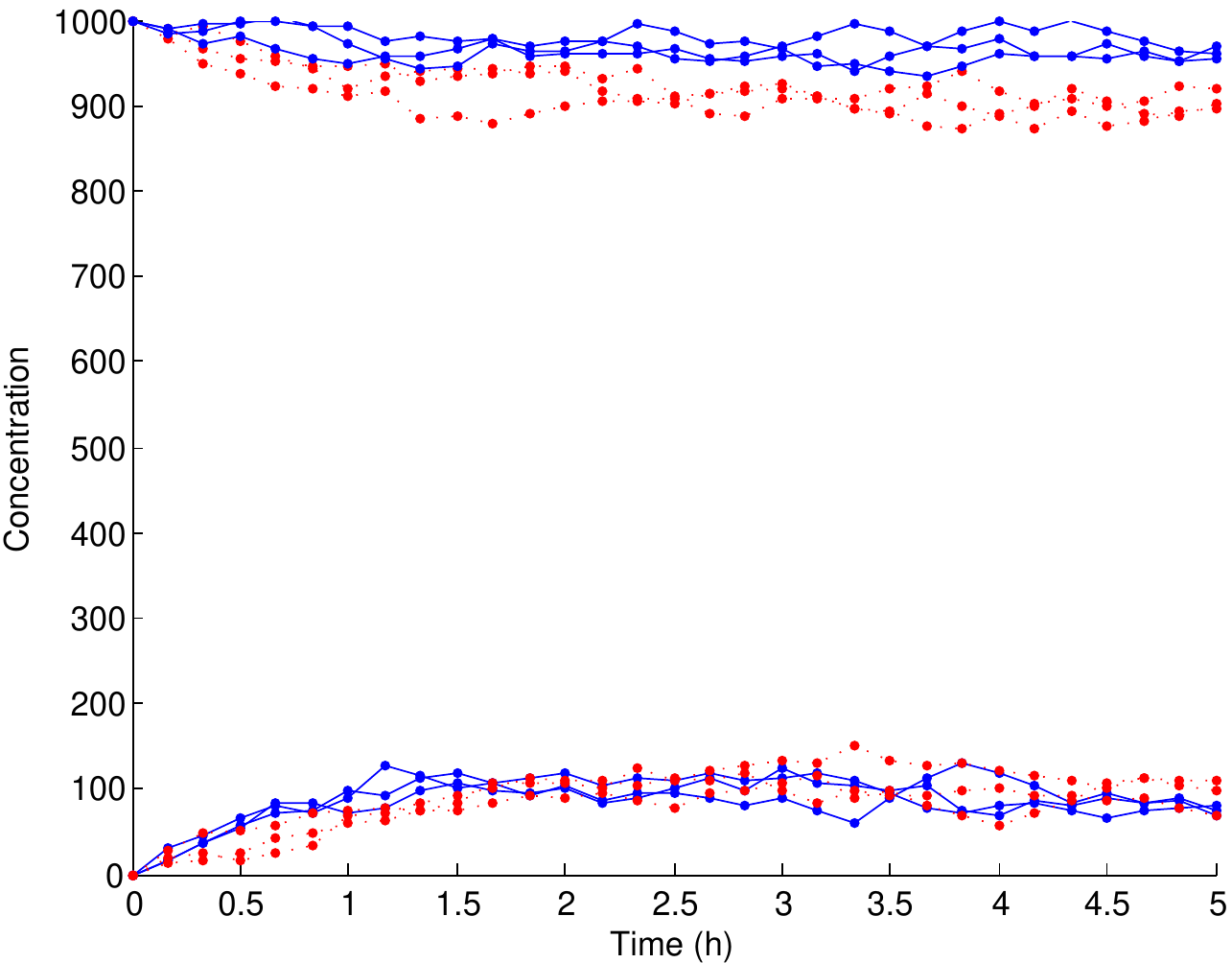}\label{fig:traj3}}

\caption{Simulated trajectories of the characterization devices (Fig. \ref{fig:pic1}) for all three promoters are shown in the first column. Red (dashed) and blue (solid) lines indicate repressor and reporter concentrations, respectively. The duration of the simulated experiment is 3 h where data is collected every 10 min (data points are shown as dots). The characterization data is used to compute the piecewise constant (second column) and piecewise linear (third column) ranges of expression rates. Simulated trajectories of the three toggle switch device models are shown in the fourth column. For devices 1,2 and 3, red (dashed) and blue (solid) lines indicates the concentrations of proteins $x_{g_1}$ and $x_{g_2}$, $x_{g_1}$ and $x_{g_3}$, and $x_{g_2}$ and $x_{g_3}$, respectively. To test the two stable equilibria for each device, trajectories are initialized at low concentrations of one protein and high concentrations of the other and vice versa. Simulations confirm the analysis results, indicating that only device 3 is bistable where, at the two equilibria, $x_{g_2}$ is more than twice greater than $x_{g_3}$ and vice versa..}\label{fig:results}
\end{figure*}

To characterize the promoters in the library, we need to obtain experimental data of their expression rates at different repressor concentrations as described in Sec. \ref{sec:problem} and \ref{sec:identification}. We follow the strategy from~\cite{Rosenfeld2005} where a characterization device similar to the one from Fig. \ref{fig:pic1} is used. It consists of an arbitrary reporter protein that is expressed from the regulated promoter to be characterized. The regulator protein is initialized at high concentration but is not expressed\footnote{experimentally, this is accomplished by expressing the regulator from an externally controlled promoter, which is switched off} and, as a result, repressor concentration decreases over time due to degradation. By measuring both repressor and reporter concentrations simultaneously we can compute the promoter characterization data as in Eqn. (\ref{eqn:reg_data}). We use numerical simulation of stochastic differential equations to generate a number of trajectories for each characterization device {\em in lieu} of single cell experimental measurements (several sample trajectories for all three promoters are shown in the first column of Fig. \ref{fig:results}). The rates of expression from each promoter and a piecewise constant and piecewise linear ranges are computed from this characterization data as described in Sec. \ref{sec:identification} and are shown in the second and third column of Fig. \ref{fig:results}, respectively.

We consider all ``toggle switch" devices with topology as in Fig. \ref{fig:toggle} that can be constructed from the set of available parts. For device 1, gene $g_2$ is expressed from promoter $p_1$ and gene $g_1$ is expressed from promoter $p_2$. For device 2, $g_3$ is expressed from $p_1$ and $g_1$ is expressed from $p_3$. For device 3, $g_2$ is expressed from $p_3$ and $g_3$ is expressed from $p_2$. For each device, we consider specifications $\phi_1 = \diamondsuit \square \pi_1$ and $\phi_2 = \diamondsuit \square \pi_2$ where $\pi_1 := x_{g_i} \geq 2x_{g_j}$ and $\pi_2 := 2x_{g_i} \leq x_{g_j}$. In other words, specification $\phi_1$ indicates that eventually and for all future times the concentration of protein $x_{g_i}$ is at least twice greater than that of protein $x_{g_j}$, while $\phi_2$ indicates the opposite. We seek a bistable device satisfying both specifications from different initial conditions.

Analysis using the procedure described in Sec. \ref{sec:analysis} leads to the computation of the relative volumes of the satisfying and violating regions for all three devices for each of the two specifications. Results for a model constructed using the piecewise linear procedure from Sec. \ref{sec:identification} are presented in Table \ref{table:results} (results obtained with the piecewise constant procedure are given in parentheses). Analysis indicates that only device 3 is bistable, which is confirmed through simulations of the stochastic differential equation models of all three devices (fourth column in Fig. \ref{fig:results}).

\section{CONCLUSION}\label{sec:conclusion}

In this paper, we presented an automated procedure for the design of functionally correct synthetic gene networks from parts. We formalized high level specifications of required device behavior as temporal logic formulas over linear inequalities in protein concentrations. We developed a procedure for the construction of device models from experimental data characterizing the different parts the devices were composed of. The identified models were related to PWA systems but allowed expression rates from promoters to vary in certain ranges and could capture all experimental observations. This model structure also allowed us to construct finite quotients through polyhedral operations. Such quotients could then be analyzed using methods inspired by model checking to compute a range of initial conditions from which all trajectories of the device model were guaranteed to satisfy (or violate) the specification. The relative sizes of those regions provided information about the correctness of a device design with respect to the specification. Our procedure could test individual, user-specified device designs or automatically search for correct devices by exploring the design space of devices constructed from a set of parts. Future research directions involve decreasing the conservatism of the method by quantifying the ``likelihood" of different parameters and applying it to real experimental studies.

\begin{table}[t]\caption{}
\vspace{-0.1in}
\begin{tabular}{|c|c|c|c|c|}
\hline	
spec. & \multicolumn{2}{|c|}{$\phi_1 = \diamondsuit \square \pi_1$} & \multicolumn{2}{|c|}{$\phi_2 = \diamondsuit \square \pi_2$} \\
\hline	
device &   satisfying & violating & satisfying & violating \\
\hline	
1 & 0\%(0\%)  & 46.9\%(44.7\%) & 35.1\%(29.2\%)  & 0\%(0\%)  \\
\hline
2 & 0\%(0\%)  & 88.8\%(85\%) & 88.8\%(25.2\%)  & 0\%(0\%)  \\
\hline
3& 8.4\%(7.4\%)  & 58.4\%(47.2\%)& 26.7\%(20.7\%)  & 8.4\%(7.4\%) \\
\hline
\end{tabular}\label{table:results}
\vspace{-0.25in}
\end{table}

\bibliographystyle{abbrv}
\bibliography{upPWA}

\begin{thebibliography}{10}

\bibitem{MITparts}
{Registry of Standard Biological Parts}.
\newblock http://partsregistry.org/.

\bibitem{Antoniotti03}
M.~Antoniotti, F.~Park, A.~Policriti, N.~Ugel, and B.~Mishra.
\newblock {Foundations of a query and simulation system for the modeling of
  biochemical and biological processes}.
\newblock In {\em Proc. of the Pacific Symposium of Biocomputing (PSB’03)},
  volume~58, pages 116--127, 2003.

\bibitem{Batt08}
G.~Batt, C.~Belta, and R.~Weiss.
\newblock {Temporal Logic Analysis of Gene Networks Under Parameter
  Uncertainty}.
\newblock {\em IEEE Transactions on Automatic Control}, 53(Special
  Issue):215--229, 2008.

\bibitem{Batt2005}
G.~Batt, D.~Ropers, H.~de~Jong, J.~Geiselmann, R.~Mateescu, M.~Page, and
  D.~Schneider.
\newblock {Validation of qualitative models of genetic regulatory networks by
  model checking: analysis of the nutritional stress response in Escherichia
  coli.}
\newblock {\em Bioinformatics (Oxford, England)}, 21 Suppl 1:i19--28, June
  2005.

\bibitem{Batt07}
G.~Batt, B.~Yordanov, R.~Weiss, and C.~Belta.
\newblock {Robustness analysis and tuning of synthetic gene networks.}
\newblock {\em Bioinformatics (Oxford, England)}, 23(18):2415--22, Sept. 2007.

\bibitem{Cai2009}
Y.~Cai, M.~W. Lux, L.~Adam, and J.~Peccoud.
\newblock {Modeling structure-function relationships in synthetic DNA sequences
  using attribute grammars.}
\newblock {\em PLoS computational biology}, 5(10):e1000529, Oct. 2009.

\bibitem{Canton2008}
B.~Canton, A.~Labno, and D.~Endy.
\newblock {Refinement and standardization of synthetic biological parts and
  devices.}
\newblock {\em Nature biotechnology}, 26(7):787--93, July 2008.

\bibitem{Clarke99}
E.~M. Clarke.
\newblock {\em {Model Checking}}.
\newblock MIT Press, 1999.

\bibitem{deJong02}
H.~de~Jong.
\newblock {Modeling and Simulation of Genetic Regulatory Systems: A Literature
  Review}.
\newblock {\em Journal of Computational Biology}, 9(1):67--103, 2002.

\bibitem{deJong03}
H.~de~Jong.
\newblock {Genetic Network Analyzer: qualitative simulation of genetic
  regulatory networks}.
\newblock {\em Bioinformatics}, 19(3):336--344, 2003.

\bibitem{Densmore2009}
D.~Densmore, A.~{Van Devender}, M.~Johnson, and N.~Sritanyaratana.
\newblock {A platform-based design environment for synthetic biological
  systems}.
\newblock In {\em The Fifth Richard Tapia Celebration of Diversity in Computing
  Conference on Intellect, Initiatives, Insight, and Innovations - TAPIA '09},
  page~24, New York, New York, USA, 2009. ACM Press.

\bibitem{Drulhe08}
S.~Drulhe, G.~Ferrari-Trecate, and H.~de~Jong.
\newblock {The Switching Threshold Reconstruction Problem for Piecewise-Affine
  Models of Genetic Regulatory Networks}.
\newblock {\em IEEE Transactions on Automatic Control}, 53(Special
  Issue):153--165, 2008.

\bibitem{Gardner2000}
T.~Gardner, C.~Cantor, and J.~Collins.
\newblock {Construction of a genetic toggle switch in Escherichia coli}.
\newblock {\em Nature}, 403:339--342, 2000.

\bibitem{Gasteiger2005}
E.~Gasteiger, C.~Hoogland, A.~Gattiker, S.~Duvaud, M.~R. Wilkins, R.~D. Appel,
  and A.~Bairoch.
\newblock {Protein Identification and Analysis Tools on the ExPASy Server}.
\newblock In J.~M. Walker, editor, {\em The Proteomics Protocols Handbook},
  pages 571--607. Humana Press, 2005.

\bibitem{Hill2008}
A.~D. Hill, J.~R. Tomshine, E.~M.~B. Weeding, V.~Sotiropoulos, and Y.~N.
  Kaznessis.
\newblock {SynBioSS: the synthetic biology modeling suite.}
\newblock {\em Bioinformatics (Oxford, England)}, 24(21):2551--3, Nov. 2008.

\bibitem{Juloski05}
A.~L. Juloski, W.~Heemels, G.~Ferrari-Trecate, R.~Vidal, S.~Paoletti, and
  J.~H.~G. Niessen.
\newblock {Comparison of Four Procedures for the Identification of Hybrid
  Systems}.
\newblock In M.~Morari and L.~Thiele, editors, {\em Hybrid Systems: Computation
  and Control}, volume 3414 of {\em Lecture Notes in Computer Science}, pages
  354--369. Springer Berlin / Heidelberg, 2005.

\bibitem{Knight2003}
T.~Knight.
\newblock {Idempotent Vector Design for Standard Assembly of Biobricks Standard
  Biobrick Sequence Interface}, 2003.

\bibitem{Lin1992}
J.~N. Lin and R.~Unbehauen.
\newblock {Canonical piecewise-linear approximations}.
\newblock {\em IEEE Transactions on Circuits and Systems I: Fundamental Theory
  and Applications}, 39(8):697--699, 1992.

\bibitem{Little2010}
S.~Little, D.~Walter, K.~Jones, C.~Myers, and A.~Sen.
\newblock {Analog/Mixed-Signal Circuit Verification Using Models Generated from
  Simulation Traces}.
\newblock {\em IJFCS}, 21(02):191, 2010.

\bibitem{Milner1989}
R.~Milner.
\newblock {Communication and concurrency}.
\newblock {\em Prentice-Hall}, 1989.

\bibitem{Pedersen09}
M.~Pedersen and A.~Phillips.
\newblock {Towards programming languages for genetic engineering of living
  cells}.
\newblock {\em Journal of The Royal Society Interface}, 6(Suppl 4):S437--S450,
  2009.

\bibitem{Purnick2009}
P.~E.~M. Purnick and R.~Weiss.
\newblock {The second wave of synthetic biology: from modules to systems.}
\newblock {\em Nature reviews. Molecular cell biology}, 10(6):410--22, June
  2009.

\bibitem{Rodrigo2007}
G.~Rodrigo, J.~Carrera, and A.~Jaramillo.
\newblock {Asmparts: assembly of biological model parts.}
\newblock {\em Systems and synthetic biology}, 1(4):167--70, Dec. 2007.

\bibitem{Rosenfeld2005}
N.~Rosenfeld, J.~W. Young, U.~Alon, P.~S. Swain, and M.~B. Elowitz.
\newblock {Gene regulation at the single-cell level.}
\newblock {\em Science (New York, N.Y.)}, 307(5717):1962--5, Mar. 2005.

\bibitem{Sontag81}
E.~Sontag.
\newblock {Nonlinear regulation: The piecewise linear approach}.
\newblock {\em IEEE Transactions on Automatic Control}, 26(2):346--358, Apr.
  1981.

\bibitem{Yordanov2010}
B.~Yordanov and C.~Belta.
\newblock {Formal Analysis of Discrete-Time Piecewise Affine Systems}.
\newblock {\em IEEE Transactions on Automatic Control}, 55(12):2834--2840, Dec.
  2010.

\end{thebibliography}

\newpage
\appendix

Given a convex region $\mathcal{X}_l$, let $x \in \mathcal{X}_l$. Given vertices $v_i \in \mathcal{V}(\mathcal{X}_l)$, $i=1,\ldots,M$, $M = |\mathcal{V}(\mathcal{X}_l)|$ we have $x = \sum_{i=1}^{M}\lambda_iv_i$ for some $\lambda_i$ such that for all $i=1,\ldots,M$, $0 \leq \lambda_i \leq 1$ and $\sum_{i=1}^{M}\lambda_i = 1$. From Eqns. (\ref{eqn:range_comput_linear}) and (\ref{eqn:overall_range}) it follows that
\begin{equation}
B(x) =  B(\sum_{i=1}^{M}\lambda_iv_i) = \sum_{i=1}^{M}\lambda_iB(v_i). \nonumber
\end{equation}
From Eqn. (\ref{eqn:system_multiaffine}) we have
\begin{eqnarray}
Post(x) & \in & Ax+B(x) = \nonumber \\
        & = & A\sum_{i=1}^{M}\lambda_iv_i+B(\sum_{i=1}^{M}\lambda_iv_i) = \nonumber \\
        & = & \sum_{i=1}^{M}\lambda_i(Av_i+B(v_i)) \Rightarrow \nonumber \\
Post(x) &\in& hull(\{Av + B(v)\;|\;v \in \mathcal{V}(\mathcal{X}_l)\}). \nonumber
\end{eqnarray}
Similarly, let $x' \in hull(\{Av + B(v)\;|\;v \in \mathcal{V}(\mathcal{X}_l)\})$. Then, for some $\mu_i$ such that for all $i=1,\ldots,M$, $0 \leq \mu_i \leq 1$ and $\sum_{i=1}^{M}\mu_i = 1$
\begin{eqnarray}
  x'  & \in &  \sum_{i=1}^{M}\mu_i(Av_i+B(v_i)) \Rightarrow \nonumber \\
  x'  &\in & Ax+B(x)=Post(x) \mbox{ where } \nonumber \\
  x   & = & \sum_{i=1}^{M}\mu_iv_i \in \mathcal{X}_l\nonumber
\end{eqnarray}

\end{document}